\documentclass{aa}
\usepackage{graphicx}
\usepackage{txfonts}
\usepackage[colorlinks,linkcolor=blue,urlcolor=cyan,citecolor=blue]{hyperref}
\usepackage{amsmath}

\graphicspath{{fig/}}  

\newcommand{\ha}{\hbox{H$\alpha$}}
\newcommand{\hb}{\hbox{H$\beta$}}
\newcommand{\hg}{\hbox{H$\gamma$}}
\newcommand{\hd}{\hbox{H$\delta$}}
\newcommand{\he}{\hbox{H$\epsilon$}}

\newcommand{\hii}{\hbox{H\,{\sc ii}}}
\newcommand{\hei}{\hbox{He\,{\sc i}}}

\newcommand{\gsim}{\lower.5ex\hbox{$\; \buildrel > \over \sim \;$}}
\newcommand{\lsim}{\lower.5ex\hbox{$\; \buildrel < \over \sim \;$}}

\newcommand{\oiii}{\hbox{[O\,{\sc iii}]}}
\newcommand{\nii}{\hbox{[N\,{\sc ii}]}}
\newcommand{\sii}{\hbox{[S\,{\sc ii}]}}

\begin{document}

   \title{Nebular Dust Attenuation with the Balmer and Paschen Lines based on the MaNGA Survey}
   \author{Zesen Lin
          \inst{1,2}
          \and
          Renbin Yan\inst{1,2}}

   \institute{$^1$ Department of Physics, The Chinese University of Hong Kong, Shatin, N.T., Hong Kong S.A.R., China\\
   $^2$ CUHK Shenzhen Research Institute, No.10, 2nd Yuexing Road, Nanshan, Shenzhen, China\\
      \email{zslin@cuhk.edu.hk, rbyan@phy.cuhk.edu.hk}
      }


  \abstract{Dust attenuations observed by stars and ionized gas are not necessarily the same. The lack of observational constraints on the nebular dust attenuation curve leaves a large uncertainty when correcting nebular dust attenuation with stellar continuum-based attenuation curves. Making use of the DAP catalogs of the MaNGA survey, we investigate the nebular dust attenuation of \hii\ regions traced by the Balmer and Paschen lines. Based on a simple simulation, we find that star-forming regions on kpc scales favor the classic foreground screen dust model rather than the uniform mixture model. We propose a novel approach to fit the dust attenuation curve using the emission-line fluxes directly. For strong hydrogen recombination lines (e.g., \hg, \hd, and \he), the slopes of the nebular attenuation curve can be well determined and are found to be in good agreement with the Fitzpatrick Milky Way extinction curve with an accuracy of $\lesssim 4\%$ in terms of the correction factor. However, severe contaminations/systematic uncertainties prevent us from obtaining reasonable values of the slopes for weak recombination lines (e.g., the high-order Balmer lines or the Paschen lines). We discuss how the choice of emission line measurement methods affects the results. Our results demonstrate the difficulty of deriving an average nebular dust attenuation curve given the current ground-based emission-line measurements.}


   \titlerunning{Nebular Dust Attenuation based on the MaNGA Survey}
   \authorrunning{Lin \& Yan}
   \maketitle
\section{Introduction}
\label{sec: intro}

Dust in the interstellar medium (ISM) prevents us from measuring directly the intrinsic emission of galaxies, from either stars or ionized gas. Observationally, about one-half of ultraviolet and optical emission is absorbed by dust and re-emitted in infrared wavelengths \citep{Dole2006}, 95\% of which is contributed by star-forming galaxies (SFGs; \citealt{Viero2013}). Such dust effects need to be accounted for when accurately estimating some galactic properties, especially for star formation rate (SFR; e.g., \citealt{Kennicutt1998, Kennicutt2012}).


Correcting for dust reddening requires knowledge about the wavelength dependence of dust scattering and absorption, which is known as dust extinction or attenuation curve (see \citealt{Calzetti2001} and \citealt{Salim2020} for reviews).\footnote{By definition, dust extinction describes light loss along individual sightlines, including dust absorption and scattering out of the sightline, which is typical for point sources like stars and quasars. For extended sources, the geometry between the emitters and dust should be considered, and thus additional effects (such as dust scattering into the sightline and varying dust column densities along different sightlines) should be taken into account, which is referred to as dust attenuation.}
Lots of efforts have been made to derive the dust extinction/attenuation curve from local (e.g., \citealt{Cardelli1989, Calzetti2000, Battisti2016, Battisti2017}) to high-redshift (e.g., \citealt{Reddy2015, Scoville2015, Battisti2022}) Universe and from integrated light of galaxies to sub-galactic resolved \hii\ regions (e.g., \citealt{Teklu2020, Liniu2020, Calzetti2021}). However, most of the studies in the literature only focused on the dust attenuation of the stellar continuum (i.e., emission from stellar objects). On the other hand, emission lines arising from ionized gas are important tracers of physical properties of galaxies, such as SFR \citep{Kennicutt1998, Kennicutt2012}, gas-phase metallicity, ionization parameter, and so on \citep{Kewley2019a}. The dust attenuation effect on emission lines might not be the same as that for the stellar continuum.

Previous studies have demonstrated that nebular emission lines tend to suffer more dust attenuation compared to the stellar continuum on average (e.g., \citealt{Calzetti1997, Kreckel2013, Price2014, Zahid2017, Qin2019a, Lin2020}), mainly attributed to the star-to-dust geometry in galaxies \citep{Calzetti1994, Charlot2000, Wild2011, Chevallard2013}. Additionally, the size distributions of dust grains in diffuse ISM and star-forming regions might be different due to the different local environments \citep{Relano2016, Relano2018, Paradis2023}. Given that both the size distribution of dust grains and the dust geometry are the main factors affecting the dust attenuation curve \citep{Weingartner2001, Draine2003, Seon2016, Narayanan2018}, it is not necessarily the case that emission lines from ionized gas in star-forming regions share the same dust attenuation curve as the stellar light, while the latter is dominated by emission from intermediate or old stellar populations outside the \hii\ regions.

To correct the stellar continuum of SFGs for dust attenuation, the \cite{Calzetti2000} curve derived from local starburst galaxies is always adopted. This usage is reasonable on average at most of the optical wavelengths and is supported by recent studies on both galactic \citep{Battisti2016} and sub-galactic \citep{Teklu2020, Liniu2020} scales. While for emission lines arising from \hii\ regions, the Milky Way extinction curves (e.g., \citealt{Cardelli1989,Fitzpatrick1999}) are also considered (e.g., \citealt{Kewley2008,LinZ2017,Ji2022a,Ji2023}). However, this convention seems to have few observational foundations. Furthermore, \cite{Ji2023} found that emission lines arising from different ionized states might follow different nebular attenuation curves, revealing the complexity of the nebular dust attenuation. Therefore, it is important to have an observationally constrained nebular dust attenuation curve that is designed for correcting dust reddening for emission lines.

Since the intrinsic flux ratios of hydrogen recombination lines of star-forming regions can be predicted assuming a reasonable electron temperature and density \citep{Osterbrock1989, Storey1995}, the Balmer decrement is widely used to estimate nebular dust attenuation giving a dust extinction/attenuation curve. Moreover, with the typical ISM condition of local star-forming regions/galaxies, the intrinsic line ratios between hydrogen recombination lines are nearly independent of the recombination models \citep{Osterbrock1989}. As such, a simple assumption of constant intrinsic recombination line ratios should be a good approximation for most local star-forming regions/galaxies studied in the literature, which also enables statistical studies of nebular attenuation. When several recombination lines are available simultaneously, deriving a nebular attenuation curve becomes feasible (e.g., \citealt{Reddy2020, Rezaee2021}). However, the Balmer series only extends up to a wavelength of 6563 \AA, beyond which the wavelength regime is still important, especially in the era of JWST that provides a wealth of near-infrared (NIR) observations of high-redshift galaxies (e.g., \citealt{Calabro2023, Reddy2023}). Despite the observational difficulty, the Paschen series also should be considered in the derivation process since these lines allow us to extend the wavelength coverage to NIR wavelengths.

Using the Paschen lines to constrain dust attenuation in \hii\ regions or SFGs has been investigated for decades (e.g., \citealt{Greve1994, Petersen1996, Calabro2018, Cleri2022, Reddy2023}). For instance, \cite{Greve1994} and \cite{Greve2010} studied dust extinction in galactic nebulae using the Balmer-Paschen line ratios between two lines with common upper atomic level (e.g., the \hg-P$\gamma$ ratio), which are known to have minimal dependence on recombination models (i.e., being nearly independent of electron temperature and density) and thus trace dust reddening precisely \citep{Osterbrock1989}. Recently, \cite{Reddy2023} used spectra observed from the JWST/NIRSpec to derive dust attenuation from the Paschen lines for galaxies at $z=1.0-3.1$ and found higher SFR when the Paschen lines are used.

Among the efforts of deriving nebular extinction/attenuation curve of \hii\ regions or SFGs, some of them only considered the blue part of the optical emission \citep{Reddy2020, Rezaee2021}, namely covering a very narrow spectral window from \he\ to \ha\ (i.e., from $\sim 3900$ to $\sim 6600$ \AA). \cite{Bautista1995} derived an empirical nebular extinction curve using the Balmer, Paschen, and Brackett recombination series observed in the Orion nebula. Utilizing HST grism spectra, \cite{Prescott2022} combined P$\beta$ with the Balmer lines to constrain the slope of the curve but found very large scatters. The NIR regime of the nebular attenuation curve is much less explored so far.

On the other hand, nearly all studies attempting to derive dust attenuation assumed a foreground dust screen model, in which the Paschen and Balmer lines are affected by the same amount of dust. This assumption was only examined by several studies (e.g., \citealt{Hulst1988, Calzetti1996, Petersen1997, Tomivcic2017}). Based on bright \hii\ regions in nearby galaxies, \cite{Petersen1997} found that the dust nucleus model also fits the observed data as well as the foreground screen model does. By studying the relation between nebular attenuation traced by the \ha/\hb\ ratio and dust surface density in several local galaxies, \cite{Kreckel2013} found that the dust follows a distribution between the screen and mixed models. However, using a similar methodology, \cite{Tomivcic2017} revealed that the dust in M31 closely follows a foreground screen model. This appears to conflict with \cite{Kreckel2013} but can be explained by the mixture between gas obscured by dust screen and additional ionized gas suffering little attenuation (e.g., diffuse ionized gas; DIG) in the \cite{Kreckel2013} sample. Moreover, other options of dust geometry models are preferred in some extreme galaxies \citep{Calabro2018} or regions on small physical scales \citep{Liu2013b}. It is noteworthy that all the above dust geometry studies are based on small galaxy samples or even a single galaxy. Therefore, we should examine the dust geometry for \hii\ regions with a large galaxy sample.

Given the lack of a nebular attenuation curve covering a wide wavelength range, this work aims to present a preliminary study of testing the dust-to-gas geometry and deriving the nebular attenuation at several wavelengths of the Balmer and Paschen lines. The layout of this paper is as follows. In Section \ref{sec: data}, we briefly introduce the data and sample selections. In Section \ref{sec: geometry}, we present an analysis focused on the emitter-to-dust geometry on the kpc scales observed by the hydrogen recombination lines. A novel fitting method to constrain the slope of the nebular dust attenuation curve is introduced in Section \ref{sec: fit_emfluxes}. Finally, we summarize in Section \ref{sec: summary}. Throughout this paper, a flat $\Lambda$CDM cosmology with $H_0=70~\mathrm{km~s^{-1}~Mpc^{-1}}$, $\Omega_{\Lambda}=0.7$, $\Omega_m=0.3$ is adopted.


\section{Data and Sample Selection}
\label{sec: data}

\subsection{MaNGA Survey and DAP Products}
\label{subsec: MaNGA_DAP}

In this work, we make use of the integral-field spectroscopy (IFS) data from the Mapping Nearby Galaxy at Apache Point Observatory (MaNGA) survey \citep{Bundy2015, Drory2015, Yan2016}. As one of the major programs of the Sloan Digital Sky Survey IV (SDSS-IV; \citealt{Blanton2017}), the MaNGA survey carried out IFS observations for $\sim 10000$ galaxies in the local Universe. The calibrated data cubes cover a spectral range of 3622--10354 \AA\ and have a spatial resolution of about 2\farcs54, enabling spatially resolved studies down to $\sim 1-2$ kpc \citep{Law2015, Law2016, Wake2017}. Relative flux calibration achieves an rms of $<5\%$ between the wavelengths that most studies are interested in \citep{Yan2016b}. The complete data, including the data cubes and the products from the Data Analysis Pipeline (DAP; \citealt{Westfall2019, Belfiore2019}), have been released as part of the SDSS DR17 \citep{Abdurrouf2022}.

The DAP provides four types of products with the combinations of different spatial binning schemes and different continuum template sets used for the stellar continuum modeling in the emission-line measurements \citep{Abdurrouf2022}. In this work, we adopt the one labeled as \texttt{HYB10-MILESHC-MASTARHC2}, which means the following practices are adopted in the analysis.
(1) Spaxels are binned to reach a signal-to-noise ratio (SNR) of $\sim 10$ using the Voronoi algorithm \citep{Cappellari2003} when performing the stellar kinematics fitting, while the emission-line measurements are done for individual spaxels. (2) The \texttt{MILES-HC} library, only covering the wavelength range of $3575-7400$ \AA, is used in the stellar kinematics fitting. (3) In the emission line measurements, the continuum is fitted with the \texttt{MASTARHC2} template set which is constructed based on a subset of the high-quality spectra from the MaNGA Stellar Library (MaStar, \citealt{Yan2019}) using a hierarchical clustering method. Benefitting from the wide wavelength coverage of the MaStar library, the \texttt{MASTARHC2} template set has a much wider wavelength coverage from 3622 to 10354 \AA, which enables stellar modeling at wavelengths longer than 7500 \AA\ and allows for a stellar absorption correction for the Paschen lines. For each type of catalog, emission-line fluxes derived from the Gaussian line-profile fitting and the zero-order moments of the continuum-subtracted spectra are given \citep{Westfall2019}.\footnote{The MaNGA DAP corrects for the Milky Way foreground extinction before performing both the stellar continuum and emission-line fittings, meaning that all emission-line measurements in the catalogs have been corrected for the Milky Way foreground dust. The correction uses the $E(B-V)$ values from the \cite{Schlegel1998} maps and adopts the Milky Way extinction curve of \cite{ODonnell1994}.} We only consider the Gaussian-fitted flux measurements in our following analysis as it has a much higher signal-to-noise ratio. In Appendix \ref{appendix: other_effects} we provide a brief discussion about the accuracy of the adopted emission-line measurements, especially for the \ha\ and \hb\ lines since we use their ratio (i.e., Balmer decrement) to trace nebular attenuation throughout this paper.

Given the wide wavelength coverage of the MaNGA spectra, the adopted DAP products provide emission-line measurements for the Balmer lines from \ha\ to H12 and for the Paschen lines from P8 to P10.\footnote{For convenience, we use ``H$n$'' (``P$n$'') to denote one Balmer (Paschen) line arising from a transition with an upper atomic level of $n$ when $n\geq$ 7 (8) (e.g., \citealt{Berg2015}). For example, \he\ will be denoted as H7 hereafter.}
However, besides the \ha\ and \hb\ lines that serve as a fiducial pair of our method (see Sections \ref{sec: geometry} and \ref{sec: fit_emfluxes}), we will only focus on two Balmer lines (\hg\ and \hd) and two Paschen lines (P8 and P10) in the main part of this paper, while other lines will be briefly discussed in Appendix \ref{appendix: res_otherlines}.

\subsection{Sample Selection}
\label{subsec: sample}

We use the \texttt{MANGA\_TARGET1} bitmask with a bit number between 1 and 12 to select galaxies from the MaNGA survey.\footnote{\url{https://www.sdss4.org/dr17/algorithms/bitmasks}}
To ensure reliable measurements of emission line fluxes, we require spaxels to have SNRs $\geq 3$ for all lines used in star-forming region selection (i.e., \ha, \hb, $\oiii~\lambda$5007, $\nii~\lambda$6584, and $\sii~\lambda\lambda$6717,6731). Furthermore, given the aim of investigating nebular attenuation with the Paschen lines, we only include galaxies with a redshift of $z\leq 0.08$ to make sure that the P8 lines are covered in the MaNGA spectra. We apply a 3D star-forming region selection scheme proposed by \cite{Ji2020}, namely
\begin{align}
    \mathrm{P1} &= 0.63\mathrm{N2} + 0.51\mathrm{S2} + 0.59\mathrm{R3},\\
    \mathrm{P2} &= -0.63\mathrm{N2} + 0.78\mathrm{S2},\\
    \mathrm{P1} &< -1.57\mathrm{P2}^2 + 0.53\mathrm{P2} - 0.48, \label{eq: SF_sel}
\end{align}
in which $\mathrm{N2}\equiv \log(\nii~\lambda 6584/\ha)$, $\mathrm{S2}\equiv \log(\sii~\lambda\lambda 6717,6732/\ha)$, and $\mathrm{R3}\equiv \log(\oiii~\lambda5007/\hb)$. This scheme results from the effort of better classifying and modeling ionization sources presented by \cite{Ji2020}, in which the authors reprojected the MaNGA spaxels in the N2--S2--R3 3D space onto a new 2D (i.e., P1--P2) plane in a way that makes both distribution surfaces of star-forming and active galactic nucleus (AGN) regions edge-on. By modeling both surfaces on the P1--P2 diagram, they further quantified the fractional AGN contribution to the \ha\ fluxes of spaxels and introduced a demarcation curve (i.e., Equation (\ref{eq: SF_sel})) to select star-forming spaxels with an AGN contribution $\lesssim$ 10\%. This criterion combines the widely-used \nii- and \sii-based Baldwin--Phillips--Terlevich (BPT) diagrams \citep{Baldwin1981, Veilleux1987} and could reconcile the classification inconsistency between those criteria based on either the \nii- or the \sii-BPT diagrams (e.g., \citealt{Kewley2001, Kauffmann2003a}). \hii\ regions at the outskirts of galaxies that might be missed by the traditional BPT criteria could also be selected with the new P1--P2 diagnostics \citep{Ji2022a}. Selecting an \hii-region sample with the \nii--\oiii\ diagram and the \cite{Kauffmann2003a} demarcation only changes a very small fraction of our sample and would not alter our results.

\begin{figure}[htb!]
    \centering
    \includegraphics[width=0.45\textwidth]{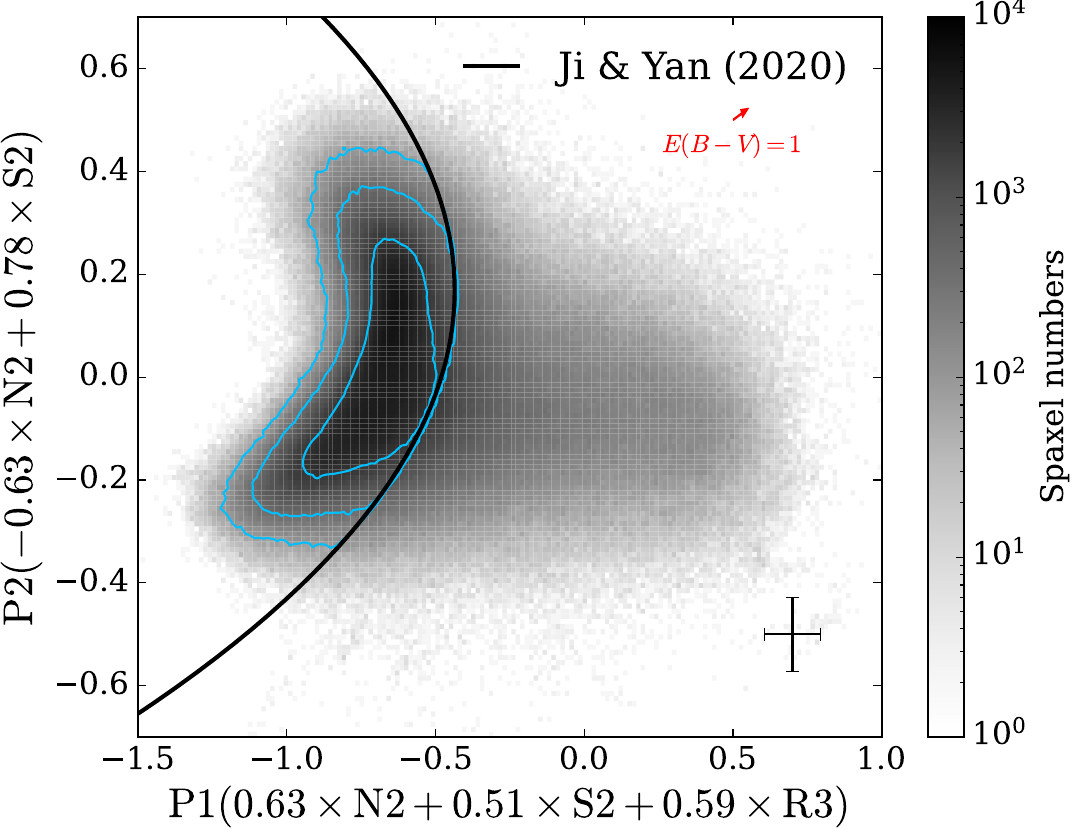}
    \caption{Distribution of the SNR-selected spaxels in the P1--P2 diagram. The black solid curve is the \hii\ region selection function given by \cite{Ji2020}. The background grayscale map shows the number density distribution of all the SNR-selected spaxels, while the cyan contours enclose 68\%, 95\%, and 99\% of the \hii\ region spaxels only, respectively. The median uncertainties of the SNR-selected sample are given in the lower-right corner, while the dust attenuation vector indicating $E(B-V)=1$ mag is shown in the upper-right corner.
    \label{fig: HII_diag}}
\end{figure}

To apply the \hii-region selection, we correct the three line ratios (i.e., N2, S2, and R3) for galaxy internal dust using the Milky Way dust extinction curve of \cite{Fitzpatrick1999} with $R_V=3.1$.
\footnote{One should note that the N2, S2, and R3 indexes used here are all insensitive to dust correction due to the very small wavelength differences between the lines involved in each index. Therefore, the choice of dust extinction/attenuation curve should not have a significant influence on the sample selection. We also plot the dust attenuation vector indicating $E(B-V)=1$ mag in Fig. \ref{fig: HII_diag} for reference.} Because this work focuses on nebular dust attenuation itself, these dust-corrected emission-line fluxes are only used in sample selection steps, i.e., \hii\ region selection here and SFR surface density selection in Section \ref{subsubsec: simu_fitting_fitsubsample}.
The above criteria result in about $3.48 \times 10^6$ spaxels from 5623 galaxies, which are taken as the main sample of this work. The projected distribution of all spaxels on the P1--P2 plane is presented in Fig. \ref{fig: HII_diag}. The star-forming population occupies the region with the largest number density on the P1--P2 plane and accounts for 78.5\% of the SNR-selected spaxels.

In this work, the SNR cut at 3 is only applied to the aforementioned emission lines that are involved in the \hii\ region selection. Although the nebular dust attenuation will be investigated at the wavelengths of other Balmer and Paschen lines, we would not put any SNR cut on these lines for the reason of reducing the bias arising from sample selection.

\section{Dust Geometry Observed by the Hydrogen Recombination Lines}
\label{sec: geometry}

\subsection{Two Typical Geometry Models}
\label{subsec: 2_models}

The geometry between the emitting sources (i.e., ionized gas in this study) and dust plays an important role in shaping the observed dust attenuation even when the underlying extinction curve is given \citep{Seon2016, Narayanan2018}. Simple analytic models related to local geometry between the emitters and dust have been widely studied \citep{Natta1984, Calzetti1994, Witt1996, Gordon1997, Witt2000}. Generally speaking, the inclusion of geometry could lead to dramatic variations in both the slope of the dust attenuation curve and the strength of the UV absorption bump at 2175 \AA\ \citep{Seon2016, Narayanan2018, Salim2020}. Observational constraints are difficult for extragalactic \hii\ regions or SFGs due to the unknown and complex intrinsic shape of the stellar continua and the underlying dust extinction curve, both of which are very difficult to constrain for extragalactic systems.

Fortunately, due to the predictable intrinsic emissivity ratio, the flux ratio between different hydrogen recombination lines becomes one available indicator of the dust geometry and has been widely used in small sample studies (e.g., \citealt{Calzetti1994, Calabro2018}) or individual galaxies (e.g., \citealt{Liu2013b}). Here we apply a similar method to the MaNGA sample and attempt to determine the general dust geometry for extragalactic \hii\ regions on kpc scales.

Taking into account the large scatter of the MaNGA data in the line ratio plots owing to the large emission-line uncertainties, we only consider two simple dust geometry models that are representative, i.e., the foreground dust screen model and the uniform mixture model. Other simple geometry models that were considered in the literature always show intermediate properties between the above two \citep{Natta1984, Calzetti1994, Liu2013b}. The former model is the simplest one in which dust is uniformly distributed and only acts as a foreground screen. The distance to the emitters is assumed to be large enough to avoid any scattering back to the sightline. Thus, all recombination lines from the same \hii\ region are attenuated by the same amount of dust. The observed flux can be described by \citep{Calzetti1994}
\begin{equation} \label{eq: fobs_screen}
f_{\rm rec, obs}=f_{\rm rec, int}\times 10^{-0.4A_{\rm rec}}=f_{\rm rec, int}\times 10^{-0.4k_{\rm rec}E(B-V)},
\end{equation}
in which $f_{\rm rec, int}$ and $f_{\rm rec, obs}$ are the intrinsic and observed fluxes of any recombination line, respectively. $A_{\rm rec}$ is the total attenuation of the dust screen at the wavelength of the recombination line and can be expressed as $A_{\rm rec}=k_{\rm rec}E(B-V)$. Here, $k_{\rm rec}$ is the total-to-selective attenuation at the wavelength of the recombination line and $E(B-V)$ is the corresponding color excess of the dust screen. Naturally, we can assume that all the recombination lines arising from the same \hii\ regions suffer the same dust attenuation, meaning that Equation (\ref{eq: fobs_screen}) for different recombination lines will have the same $E(B-V)$. By comparing Equation (\ref{eq: fobs_screen}) between different recombination lines, for example \ha\ and \hb, we can obtain
\begin{equation} \label{eq: ebv_screen1}
    E(B-V)=\frac{2.5}{k_{\rm H\beta}-k_{\rm H\alpha}}\log\left(\frac{f_{\rm H\alpha, obs}/f_{\rm H\beta, obs}}{f_{\rm H\alpha, int}/f_{\rm H\beta, int}}\right).
\end{equation}
Similarly, other recombination line ratio, say $f_{\rm X}/f_{\rm H\alpha}$ (``X'' could be any hydrogen recombination line other than \ha\ and \hb), should follow
\begin{equation} \label{eq: ebv_screen2}
    E(B-V)=\frac{2.5}{k_{\rm H\alpha}-k_{\rm X}}\log\left(\frac{f_{\rm X, obs}/f_{\rm H\alpha, obs}}{f_{\rm X, int}/f_{\rm H\alpha, int}}\right).
\end{equation}

Since $E(B-V)$ in Equations (\ref{eq: ebv_screen1}) and (\ref{eq: ebv_screen2}) are assumed to be the same, it is easy to obtain that when the foreground screen model is applied, the relation between line ratios $f_{\rm X}/f_{\rm H\alpha}$ and $f_{\rm H\alpha}/f_{\rm H\beta}$ can be written as
\begin{align} \label{eq: lineratio_screen}
    \log\left(\frac{f_{\rm X,obs}}{f_{\rm H\alpha,obs}}\right)
    =& \frac{k_{\rm H\alpha}-k_{\rm X}}{k_{\rm H\beta}-k_{\rm H\alpha}}\left[\log\left(\frac{f_{\rm H\alpha,obs}}{f_{\rm H\beta,obs}}\right) - \log\left(\frac{f_{\rm H\alpha,int}}{f_{\rm H\beta,int}}\right)\right] \nonumber\\
    & + \log\left(\frac{f_{\rm X,int}}{f_{\rm H\alpha,int}}\right).
\end{align}
Hence, in this model, the observed line ratios $\log(f_{\rm X, obs}/f_{\rm H\alpha, obs})$ and $\log(f_{\rm H\alpha, obs}/f_{\rm H\beta, obs})$ follow a linear relation with a slope of $(k_{\rm H\alpha}-k_{\rm X})/(k_{\rm H\beta}-k_{\rm H\alpha})$, which only depends on the adopted dust extinction/attenuation curve. Obviously, any variations in the slope would reflect variations in the nebular dust attenuation curve, while deviations from the assumed Case B ratios and/or any multiplicative systematic uncertainties in relative flux calibration would be manifested as an overall shift of the relation along either direction while keeping the slope unchanged.

In the uniform mixture model (also known as ``internal dust''; \citealt{Natta1984, Calzetti1994}), the emitting sources and dust are uniformly mixed, and thus scattering and differentiated attenuation should be carefully treated. The observed fluxes for one recombination line can be written as \citep{Natta1984, Calzetti1994}
\begin{equation}
    f_{\rm rec, obs}=f_{\rm rec, int}\frac{1-10^{-0.4A_{\rm rec}}}{0.4\ln 10\times A_{\rm rec}}.
\end{equation}
Detailed derivation of this relation can be found in Section 4.2 of \cite{Calzetti1994}. A simple calculation for this model could show that any attenuated line ratios would asymptote to a constant with increasing total dust attenuation $A_V$. In other words, the observed recombination lines are only contributed by emitting sources in regions with small $A_V$, while emission from ``deeper'' regions (i.e., with large $A_V$) will be fully absorbed. Due to the wavelength-dependent opacity, the escaped recombination lines at different wavelengths originate from different depths in the dust. Typical examples of this dust model include \hii\ regions on small scales \citep{Liu2013b} and the core of a starburst galaxy \citep{Calabro2018}. The feature of saturated line ratio can be easily identified in the line ratio plots (e.g., \citealt{Calzetti1994, Liu2013b}; also see the orange dotted lines in Fig. \ref{fig: lineratio}) and will help us distinguish this model from the foreground screen one.

\subsection{Comparisons between the Model-predicted and Observed Line Ratios}
\label{subsec: comp_lineratio}

\begin{figure*}[htb!]
    \centering
    \includegraphics[width=0.95\textwidth]{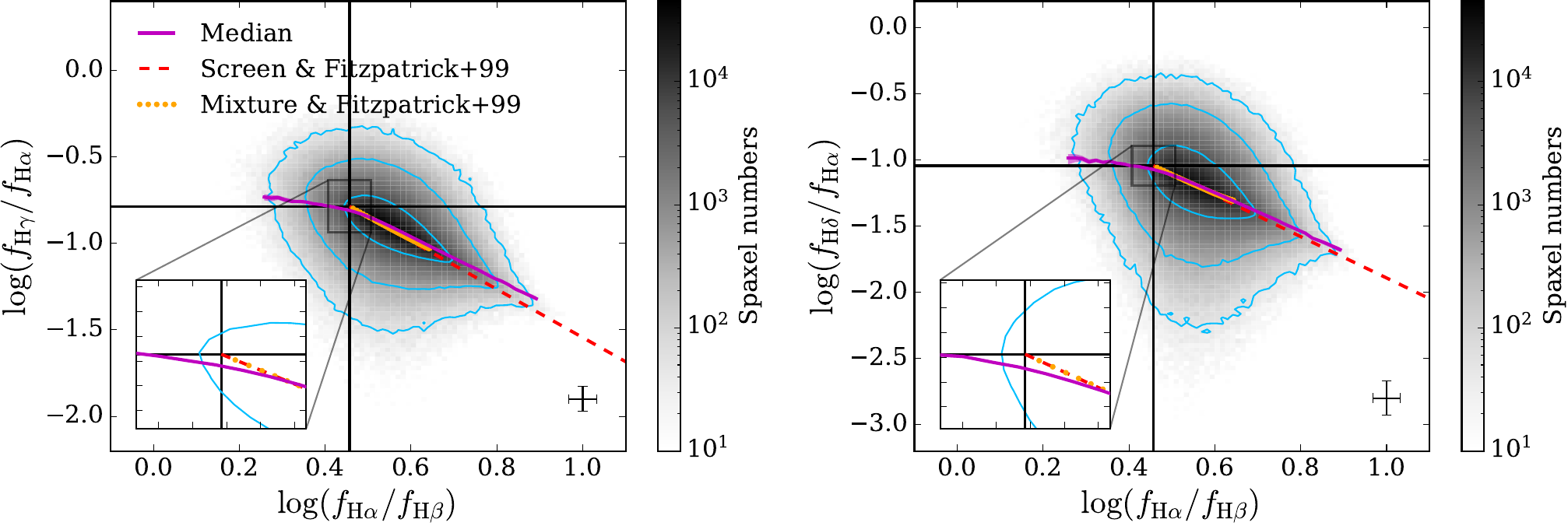}
    \includegraphics[width=0.95\textwidth]{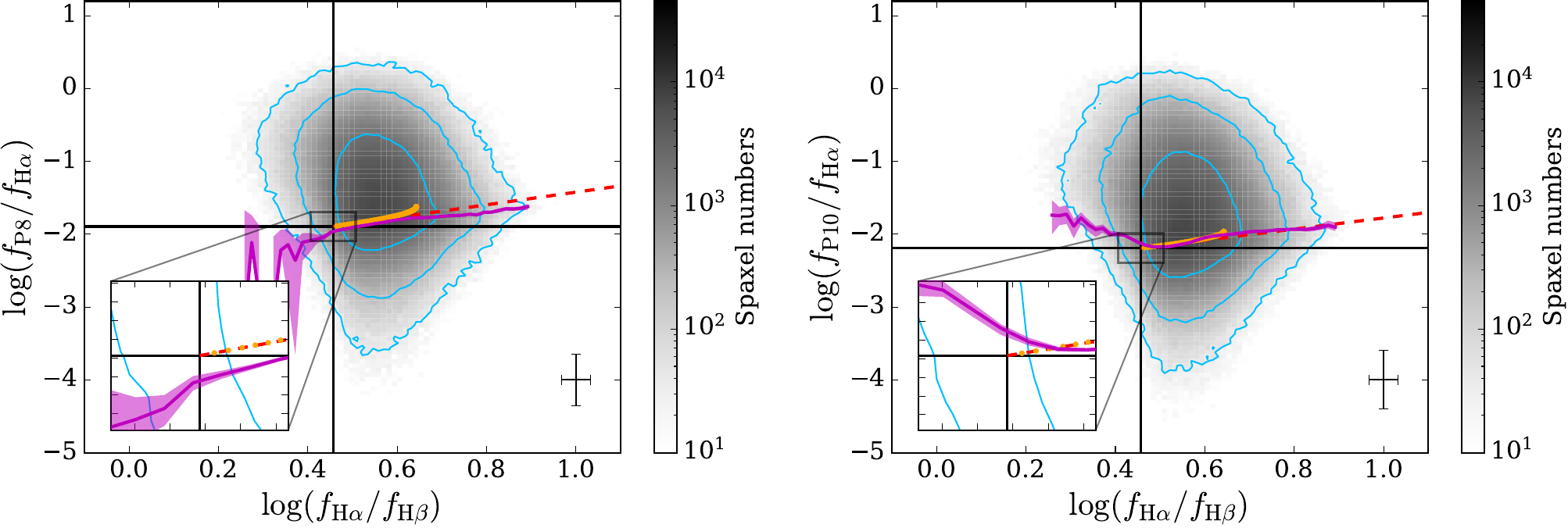}
    \caption{Comparisons between the predictions from two geometry models and the observed $f_{\rm H\alpha}/f_{\rm H\beta}$--$f_{\rm X}/f_{\rm H\alpha}$ line ratios for \hg, \hd, P8, and P10. The magenta curves and shaded regions indicate the medians and their $3\sigma$ uncertainties of the binned distributions, respectively. In each panel, the red dashed line denotes the predicted relation based on the foreground screen dust model, while the orange dotted line is the prediction based on the uniform mixture model. The \cite{Fitzpatrick1999} dust extinction curve with $R_V=3.1$ is assumed in both cases. The background gray map shows the logarithmic number density of our sample, while the cyan contours enclose 68\%, 95\%, and 99\% of the \hii\ region spaxels, respectively. The horizontal and vertical black lines indicate the intrinsic values of the corresponding ratios based on the Case B recombination. The median uncertainties of the sample are given in the lower-right corner. A close-up of the region around the cross point of the intrinsic line ratios is shown in the lower-left inset of each panel.  
    \label{fig: lineratio}}
\end{figure*}

We present the $f_{\rm H\alpha}/f_{\rm H\beta}$--$f_{\rm X}/f_{\rm H\alpha}$ line ratio diagrams ($\rm X= \hg, \hd, P8$, and P10) in Fig. \ref{fig: lineratio}. In each panel, the background grayscale map is the spaxel number density on a logarithmic scale, while the magenta curve denotes the binned medians.
The shaded regions around the median curves are their 3$\sigma$ uncertainties, which is the central 99\% percentile range of the distributions of the medians resampled via the bootstrapping method. For comparison, the predicted relations between these line ratios based on the foreground screen dust model (red dashed line) and uniform mixture model (orange dotted line) are also given by assuming the \cite{Fitzpatrick1999} dust extinction curve with $R_V=3.1$ and the Case B recombination with electron temperature $T_{\rm e}=10^4$ K and electron density $n_{\rm e}=100~\mathrm{cm^{-3}}$. The intrinsic line ratios are computed via the \texttt{PyNeb} package \citep{Luridiana2015} with the default recombination line data from \cite{Storey1995}. Note that although Fig. \ref{fig: lineratio} plots line ratios on a logarithmic scale, the median curves are computed on a linear scale. This consideration aims to avoid any bias arising from computing logarithmic line ratios for emission lines with zero fluxes since we do not apply any SNR cut on the targeted lines.

The inclusion of spaxels with low-SNR detections of emission lines leads to very large scatters in the distributions of our sample for all lines, especially for the Paschen lines. However, as expected, all the median curves generally follow a linear relation at $f_{\rm H\alpha}/f_{\rm H\beta}\gtrsim 2.86$ (i.e., the intrinsic \ha-to-\hb\ ratio expected by the Case B recombination), which are consistent with the predictions based on the foreground screen model. Both the median curves and the number density contours extend to the unphysical region with $f_{\rm H\alpha}/f_{\rm H\beta}\lesssim 2.86$, which is possible if the uncertainties are taken into account. Predictions of both dust models are generated with an $E(B-V)$ range of $0-10$ mag, which is large enough to account for the vast majority of \hii\ regions even for starburst galaxies \citep{Calabro2018}. Obviously, the major and crucial difference between the predictions of the two dust geometry models is that the foreground screen one has a much larger range in line ratios compared to the mixed mode. Due to the aforementioned line-ratio saturation effect, the uniform mixture model cannot predict a line ratio that is large enough to cover most of the observed line ratio ranges presented in Fig. \ref{fig: lineratio} when the measurement uncertainties are not considered.

Large scatters and unphysical line ratios were also reported by \cite{Prescott2022}, in which ground-based slit and space-based slitless grism spectra were combined to draw the line ratio diagram including \ha, \hb, and P$\beta$. With their small sample, the authors attributed the scatters to the differences in the spatial covering factors between the Balmer and Paschen lines and/or the slit loss in ground-based observations. Our sample is large enough to demonstrate that even though observations were performed with the same instrument and reduced in the same manner, large scatters are still observable even for the line ratio plots including the Balmer lines only (i.e., the top panels of Fig. \ref{fig: lineratio}). In Section \ref{subsec: simu_geometry} we will show that the large coverages in the line ratio diagrams can be recovered via a simple simulation, in which the uncertainties in the emission-line measurements are taken into account without the need to differentiate the coverage factors of the Balmer or Paschen lines.

Furthermore, although the median curves between the logarithmic line ratios are in good agreement with a linear relation, the slopes seem to deviate slightly from the assumed \cite{Fitzpatrick1999} values. We will return to this problem in Section \ref{sec: fit_emfluxes}. In Fig. \ref{fig: lineratio}, we also indicate the intrinsic $f_{\rm H\alpha}/f_{\rm H\beta}$ and $f_{\rm X}/f_{\rm H\alpha}$ ratios predicted by the Case B recombination with the vertical and horizontal black solid lines, respectively. As revealed in Equation (\ref{eq: lineratio_screen}), one should expect that the median curves would go through the intersection points of two black lines if the Case B recombination is valid for our sample. However, deviations from the theoretical values are observed in all cases shown in Fig. \ref{fig: lineratio}. This inconsistency is exhibited more clearly in the zoom-in plots in the insets. We will elucidate this offset in Section \ref{subsec: simu_geometry} and attribute it to the measurement uncertainties of the emission-line fluxes and/or the night-sky line residuals rather than intrinsic deviation from the Case B recombination.

\subsection{Simple Simulation for Two Geometry Models}
\label{subsec: simu_geometry}

Due to the large measurement uncertainties of line ratios, especially for the Paschen lines, the distributions of spaxels in the line ratio plots exhibit pretty large scatters around the median curve. Here we further provide a simple simulation to argue that the observed line ratio relations for most of our spaxels are better described by the foreground screen model rather than the uniform mixture model and confirm that the observed large scatters can be predominately explained by the large measurement uncertainties of emission lines.

During the simulation, we assume the \cite{Fitzpatrick1999} extinction curve with $R_V=3.1$ and the Case B recombination with $T_{\rm e}=10^4$ K and $n_{\rm e}=100~\mathrm{cm^{-3}}$. The procedure is as follows. 
\begin{itemize}
    \item For all spaxels in our sample, the nebular color excesses $E(B-V)$ and intrinsic (dust-corrected) \ha\ fluxes ($f_{\rm H\alpha, int}$) for individual spaxels are computed from the observed $\ha/\hb$ ratio by assuming the geometry model that we want to simulate.
    \item Due to the strong correlation between $f_{\rm H\alpha, int}$ and $E(B-V)$ (i.e., the strong correlation between SFR and the nebular dust attenuation; \citealt{Li2019}), the derived intrinsic \ha\ fluxes are randomly matched with an $E(B-V)$ value in a manner that forces the resulting $f_{\rm H\alpha, int, sim}$--$E(B-V)_{\rm sim}$ distribution to be similar to the observed one.
    \item The simulated intrinsic fluxes of \hb\ and other recombination lines X can be computed from $f_{\rm H\alpha,int,sim}$ under the Case B recombination assumption.
    \item Based on the targeted dust geometry model and $E(B-V)_{\rm sim}$, the simulated true fluxes after dust attenuation for each recombination line ($f_{\rm rec, sim, att}$) can be computed.
    \item By assuming that the true attenuated flux, $f_{\rm rec,sim,att}$, and their simulated uncertainties, $\sigma_{\rm rec,sim}$, follow the observed $f_{\rm rec,obs}$--$\sigma_{\rm rec,obs}$ distributions of our sample, we randomly assign a $\sigma_{\rm rec,sim}$ to each $f_{\rm rec,sim,att}$.
    \item Finally, the simulated observed flux ($f_{\rm rec, sim, obs}$) is obtained by assuming that the measurement uncertainty follows a Gaussian distribution with a mean of $f_{\rm rec, sim, att}$ and a standard deviation of $\sigma_{\rm rec, sim}$.
\end{itemize}

\begin{figure*}
    \centering
    \includegraphics[width=0.69\textwidth]{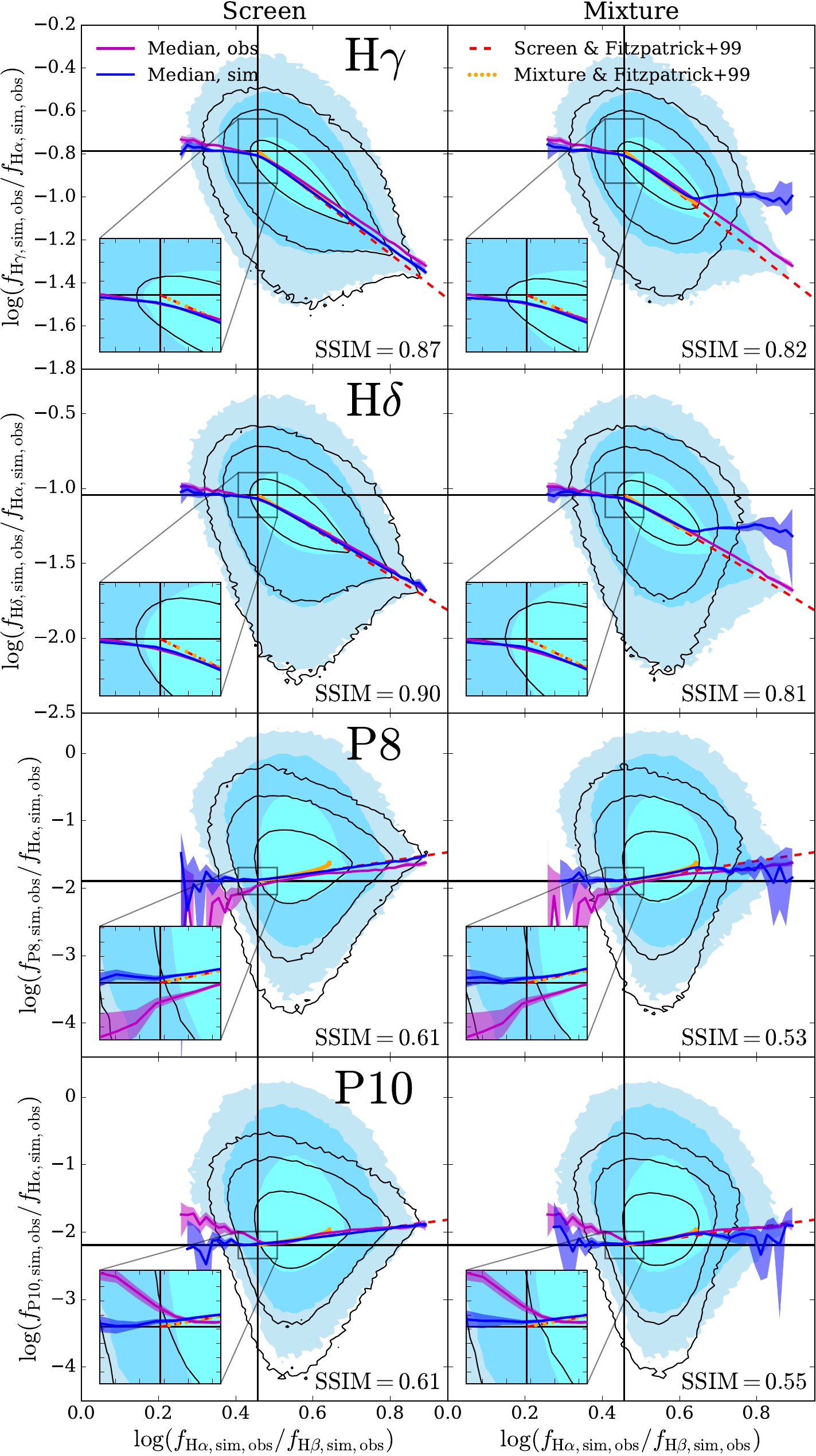}
    \caption{Comparisons of the line ratio plots between the observations and the simulated samples generated by assuming the foreground screen dust model (left) and the uniform mixture model (right). In each panel, the filled contours and black contour lines indicate the spaxel distributions of the observed and simulated samples, respectively, both of which enclose 68\%, 95\%, and 99\% of the corresponding samples from inner to outer. The magenta (blue) curve and shaded regions denote the medians and their $3\sigma$ uncertainties for the observed (simulated) sample, respectively. Regions around the cross points of the intrinsic line ratios are zoomed in and shown in the lower-left insets. The SSIM index described the similarity in density distribution between the observed and simulated data is listed in the lower-right corner of each panel. Other symbols are the same as those in Fig. \ref{fig: lineratio}.
    \label{fig: lineratio_sim}}
\end{figure*}

In short, this simulation aims to reproduce the observed $E(B-V)$--$f_{\rm H\alpha, int}$ and $f_{\rm rec,obs}$--$\sigma_{\rm rec,obs}$ relations. Although what we simulate is the $f_{\rm rec, sim, att}$--$\sigma_{\rm rec, sim}$ relation in practice, we also compare the obtained $f_{\rm rec, sim, obs}$--$\sigma_{\rm rec, sim}$ relation with the observed $f_{\rm rec, obs}$--$\sigma_{\rm rec, obs}$ and find good agreements between them for all lines. To match our selection criteria described in Section \ref{subsec: sample}, simulated spaxels with $f_{\rm H\alpha, sim, obs}/\sigma_{\rm H\alpha, sim}<3$ or $f_{\rm H\beta, sim, obs}/\sigma_{\rm H\beta, sim}<3$ are excluded from our simulated sample. Negative simulated fluxes due to the resampling from the Gaussian distribution might be obtained for some spaxels and are set to zero in the simulated catalogs. This setting is designed to mock the results of DAP, which forces the Gaussian flux to be non-negative. Thus, for non-detection, the fluxes could be zero. Given the line-ratio saturation in the uniform mixture model, about 12.6\% of our sample cannot obtain valid attenuation estimations due to their too large \ha-to-\hb\ ratios. Fig. \ref{fig: lineratio_sim} shows similar plots as Fig. \ref{fig: lineratio} but adds the results for the simulated sample generated by assuming two considered geometry models. Since the \cite{Fitzpatrick1999} extinction curve with $R_V=3.1$ is adopted in the simulation, the red dashed and orange dotted lines should be the true values of the median curves for the foreground screen model and uniform mixture model, respectively.

We first compare the simulated median curves with their true values. For the foreground screen model, the simulated median curves are in good agreement with the true values, except for the regimes below the intrinsic \ha-to-\hb\ ratio and the high-$f_{\rm H\alpha}/f_{\rm H\beta}$ ends. Given that the Case B recombination is assumed in the simulation, the true values of the simulated attenuated \ha-to-\hb\ ratios ($f_{\rm H\alpha, sim, att}/f_{\rm H\beta, sim, att}$) cannot be smaller than the intrinsic one, as shown by the red lines in Fig. \ref{fig: lineratio_sim}. However, the introduced uncertainties when mocking observations make it possible for the ``observed'' values ($f_{\rm H\alpha, sim, obs}/f_{\rm H\beta, sim, obs}$) to go below the intrinsic one. In addition, when $f_{\rm H\alpha}/f_{\rm H\beta}$ is at the intrinsic Case B value, the median simulated $f_{\rm X}/f_{\rm H\alpha}$ values would go below the intrinsic values, in exactly the same way as the observed trend for \hg\ and \hd. This is also because of uncertainties introduced. Because uncertainties scatter spaxels around in both the horizontal direction and vertical direction in these plots, many spaxels with an observed Case B $f_{\rm H\alpha}/f_{\rm H\beta}$ actually have non-zero intrinsic attenuation. Thus, their median $f_{\rm X}/f_{\rm H\alpha}$ should also display some level of attenuation. This also should be the reason for the small differences between the simulated median curves and their true values (i.e., the theoretical line) at the high-$f_{\rm H\alpha}/f_{\rm H\beta}$ ends, except that now we underestimate attenuation due to large contamination by points with intrinsically lower $f_{\rm H\alpha}/f_{\rm H\beta}$ ratios.

As for the uniform mixture model, due to the line-ratio saturation effect, $f_{\rm H\alpha, sim, att}/f_{\rm H\beta, sim, att}$ has an upper limit at $\sim 4.39$. Similarly, the introduction of uncertainties around the true values results in deviations from the orange dotted line. However, due to the narrow range of the true values in this model, the simulated median curves seem to deviate from the model predictions more or less in the whole range of $f_{\rm H\alpha, sim, att}/f_{\rm H\beta, sim, att}$, especially for the \hg\ and \hd\ lines. The uncertainties also can scatter a certain amount of spaxels to the high-$f_{\rm H\alpha, sim, obs}/f_{\rm H\beta, sim, obs}$ end, beyond the theoretical upper limit of this line ratio.

Comparing with the observations exhibited in Fig. \ref{fig: lineratio} (overplotted as filled contours and magenta curves in Fig. \ref{fig: lineratio_sim}), we find that the simulation results from the foreground screen model are generally more consistent with our sample in terms of both the median curves and the overall distributions of spaxels. Given that the physical scale sampled by the MaNGA data is about 1-2 kpc, this result is in good agreement with \cite{Liu2013b} in which the authors reported that the foreground screen model becomes consistent with observations averaged over a scale of $\gtrsim 100-200$ pc. The apparent ``non-Case B recombination'' intercepts at the intrinsic \ha-to-\hb\ ratios of \hg\ and \hd\ can be well reproduced by the simulation, indicating that these offsets can be attributed to the measurement uncertainties of emission lines only. For P8 and P10, the discrepancy between the simulated intercepts and the observed ones implies that additional factors other than the emission-line uncertainties should have nonnegligible effects on these weak line measurements, such as sky-night line residuals and/or zero fluxes from the Gaussian fitting (see Section \ref{subsec: reasons_c_deviations}).

As a sanity check, we adopt a structural similarity (SSIM) index proposed by \cite{Wang2004} to assess the similarity between the observed and simulated distributions for the two geometry models. The normalized 2D density distributions in logarithmic line ratio space (e.g., Fig. \ref{fig: lineratio} after normalization) are used to create density images for the calculation. The density images are then pixelated to allow a pixel-by-pixel calculation of the SSIM index, which is implemented by the Python function \texttt{skimage.metrics.structural\_similarity}. To describe the overall similarity between two density images, we use the value weighted by the normalized number density of the observed data. Generally speaking, the overall SSIM index is between 0 and 1 with a larger value indicating more similarity in overall structure between two density images.
The resulting SSIM indices of each line for both geometry models are given in the lower-right corner of the corresponding panels in Fig. \ref{fig: lineratio_sim}.

For \hg\ and \hd, both geometry models give large SSIM indices (i.e., $\rm SSIM >0.8$) due to the following reasons: (1) the main difference between the two geometry models are at large \ha/\hb\ values where the spaxel density already becomes smaller, and (2) the overall SSIM index is weighted by the normalized density distribution of the spaxels. The P8 and P10 lines have much smaller SSIM indices compared to \hg\ and \hd\ even for the foreground screen model, which implies that the assumptions/models involved in the current simulation are not enough to mock the emission-line measurements. In other words, we are still far from understanding the flux measurements of these weak recombination lines. A similar conclusion is also implied by the fitting results shown in Section \ref{subsec: res_min-mcmc} and is further discussed in Section \ref{subsec: reasons_c_deviations}. For all lines shown in Fig. \ref{fig: lineratio_sim}, the SSIM indices for the foreground screen model are all larger than the ones for the uniform mixture model, suggesting that the density distributions of the simulated catalogs based on the foreground screen model are more similar to the observed ones.

Overall, we conclude that for most of the spaxels in our sample, observations from all four recombination lines considered in this work favor the foreground screen dust model rather than the uniform mixture model. In the following analysis, we adopt this dust geometry as our fiducial one. The preferred dust geometry of ionized gas might depend on other physical properties. This topic is beyond the scope of this paper but is worthy of further investigation in a future paper.

\section{Deriving the Slope of the Nebular Attenuation} \label{sec: fit_emfluxes}

In this section, we derive the slope of nebular dust attenuation among hydrogen recombination lines, i.e., the slope in Equation (\ref{eq: lineratio_screen}). This would lead to the constraints on the attenuation curve at those specific wavelengths.

\subsection{Preliminary Thinking and Attempts towards the Fitting Method}
\label{subsec: pre_thinking}

Given Equation (\ref{eq: lineratio_screen}) and Fig. \ref{fig: lineratio}, the simplest and most straightforward way to constrain the slope of nebular dust attenuation is to fit the median curves displayed in Fig. \ref{fig: lineratio} to a linear function directly (e.g., \citealt{Prescott2022, Ji2023}). However, this method has several problems. First, as demonstrated in Section \ref{subsec: simu_geometry}, the measurement uncertainties have significant influences on the derived median curves, especially when the $f_{\rm H\alpha, obs}/f_{\rm H\beta, obs}$ ratios are very close to the intrinsic value. Fig. \ref{fig: lineratio_sim} also suggests that, even for the foreground screen model, the simulated median curves could have small but detectable deviations from the true values at the large Balmer decrement end. These effects prevent us from determining a reasonable \ha-to-\hb\ ratio range to perform the fitting. Second, when fitting median curves, although the uncertainties of the median values can be computed and taken into account in the fitting, the uncertainties of individual spaxels are difficult to handle and might be ignored in the fitting.

To fully consider the measurement uncertainties of individual spaxels, another method is to fit the logarithmic line ratios of all spaxels to a linear function. Although this method works for the \hg\ and \hd\ lines, it fails to provide reliable results for P8 and P10 since the best-fit slopes always have opposite signs to what we expected, and thus cannot match the observed median curves. One of the reasons for this strange issue is that, due to the faintness of the Paschen lines, the Gaussian fitting in the DAP resulted in zero flux for a large number of spaxels. Although the uncertainties are not zero for these spaxels, computing any logarithmic line ratios would remove them from our sample and inevitably bias the sample towards \hii\ regions that have brighter and positively biased Paschen lines or are significantly contaminated by night-sky line residuals.

Furthermore, we note that it is difficult to properly take into account the uncertainties for logarithmic line ratios. The general error propagation formula provides a computable method to give symmetric uncertainty that is easy to take into account in the fitting. However, this symmetric uncertainty of the logarithmic line ratio is approximately correct only when the uncertainties are very small. On the other hand, during the fitting, we always assume that the uncertainties of observables (i.e., the logarithmic line ratios in this case) follow a Gaussian distribution, which is also the requirement for the conventional definition of likelihood function or $\chi^2$ merit function. But it might be not true for the logarithmic line ratios since the uncertainties of individual emission-line fluxes are naturally assumed to be Gaussian, and thus the logarithmic line ratios should not follow a Gaussian distribution anymore. Given these reasons, we propose a new fitting approach to fit the measured emission-line fluxes directly in this work.

\subsection{A Novel Fitting Method Using Emission Line Fluxes Directly}
\label{subsec: fitting_method}

In this section, we provide a brief introduction to our fitting method. This method aims to constrain the slope of the nebular attenuation curve using the measured emission-line fluxes directly so that the uncertainties involved in the likelihood function (or the $\chi^2$ definition) could be simply described by Gaussian distributions.

From Equation (\ref{eq: lineratio_screen}), we can define a relative slope between \ha, \hb, and any other hydrogen recombination line (denoted as ``X'' hereafter) as
\begin{equation} \label{eq: def_m}
    m_{\rm X} =\frac{k_{\rm H\alpha}-k_{\rm X}}{k_{\rm H\beta}-k_{\rm H\alpha}}=\frac{A_{\rm H\alpha}-A_{\rm X}}{A_{\rm H\beta}-A_{\rm H\alpha}}.
\end{equation}
Taking the \cite{Fitzpatrick1999} Milky Way extinction curve with $R_V=3.1$ as the fiducial one, the observed slope for line X can be described as
\begin{equation} \label{eq: def_dm}
    m_{\rm X,obs} = \Delta m_{\rm X} + m_{\rm X,F99},
\end{equation}
in which $m_{\rm X,F99}$ is the slope computed from the Milky Way curve and $\Delta m_{\rm X}$ is the difference in slope between the observed and fiducial values. For a fixed $\Delta m_{\rm X}$, the nebular attenuation at the wavelength of line X can be computed
\begin{equation}
    A_{\rm X} = A_{\rm H\alpha} - (A_{\rm H\beta}-A_{\rm H\alpha}) m_{\rm X,obs}
\end{equation}
During the fitting, we assume that all spaxels in our sample follow the same nebular attenuation curve and treat $\Delta m_{\rm X}$ as a free parameter.

For each spaxel, we define a likelihood function related to line X as
\begin{align} \label{eq: def_lnLsp}
    \ln\mathcal{L}_{\rm spaxel} = -\frac{1}{2}\sum_{i=1}^{3}\frac{(f_{i,\rm obs}-f_{i,\rm pre})^2}{\sigma_{i,\rm obs}^2}, i=\ha, \hb, \mathrm{X}
\end{align}
where $f_{i,\rm pre} = f_{i,\rm int} 10^{-0.4A_{i}}$ is the predicted fluxes based on the considered dust attenuation curve and $\sigma_{i,\rm obs}$ is the uncertainty of emission-line flux. For a given attenuation curve and assuming the Case B recombination, we can compute all three predicted fluxes from the observed fluxes ($f_{\rm H\alpha, obs}$, $f_{\rm H\beta, obs}$, and $f_{\rm X, obs}$) with the knowledge of the intrinsic \ha\ flux ($f_{\rm H\alpha, int}$) and the dust attenuation at the wavelength of \ha\ ($A_{\rm H\alpha}$).\footnote{Here we do not introduce any variation in $k_{\rm H\beta}-k_{\rm H\alpha}$, thus $A_{\rm H\beta}$ can be computed from $A_{\rm H\alpha}$ with the fiducial extinction curve.} In other words, for each spaxel, we take $f_{\rm H\alpha, int}$ and $A_{\rm H\alpha}$ as free parameters to obtain $f_{i,\rm pre}$. The posterior probability is computed as
\begin{equation}
    \ln\mathcal{P}_{\rm spaxel} = \ln\mathcal{L}_{\rm spaxel} + \ln \mathcal{P}_{\rm prior},
\end{equation}
in which $\mathcal{P}_{\rm prior}$ is the prior probability and limits the valid range of the two parameters, i.e.,
\begin{equation}
    \mathcal{P}_{\rm prior}=\left\{
        \begin{aligned}
            &1, ~f_{\rm H\alpha, int}\geq0 ~ \& ~ A_{\rm H\alpha}\geq 0\\
            &0, ~\mathrm{otherwise}.
        \end{aligned}
    \right.
\end{equation}
For a given $\Delta m_{\rm X}$, $\ln\mathcal{P}_{\rm spaxel}$ is maximized respectively for each spaxel by finding the best combination of $f_{\rm H\alpha, int}$ and $A_{\rm H\alpha}$. Then we sum up the best-fit $\ln\mathcal{P}_{\rm spaxel}$ of individual spaxels to get the posterior probability of the whole sample for that $\Delta m_{\rm X}$, i.e.,
\begin{equation}
    \ln\mathcal{P}_{\rm sample} = \sum_{j=1}^{N_{\rm spaxel}} \ln\mathcal{P}_{j,\rm spaxel}.
\end{equation}
Then we search the best $\Delta m_{\rm X}$ that maximize $\ln\mathcal{P}_{\rm sample}$.

Additionally, we manually introduce a constant scaling factor $c$ to the predicted flux $f_{\rm X, pre}$ for each targeted recombination line, namely, the predicted flux is rewritten as
\begin{equation}
    f_{\rm X, pre} = c \times f_{\rm X, int} 10^{-0.4A_{\rm X}}.
\end{equation}
This constant is invariant for all spaxels and could account for any systematic effects such as deviations from the Case B recombination and measurement systematics arising from the stellar continuum subtraction or night-sky line residuals. Moreover, a multiplicative systematic error in flux calibration would result in a different $c$ but would not change $\Delta m_{\rm X}$. We will show in Sections \ref{subsec: res_min-mcmc} and \ref{subsec: reasons_c_deviations} that this constant is necessary for the fitting, especially for weak lines like P10. To avoid predicting negative fluxes, we adopt $\lg c$ as a free parameter in practice.

In short, the fitting procedure can be summarized as follows. We set $\Delta m_{\rm X}$ and $c$ as free parameters for the whole sample. Given one ($\Delta m_{\rm X}$, $c$) pair, $f_{i,\rm pre}$ and thus $\ln\mathcal{L}_{\rm spaxel}$ can be derived by treating $f_{\rm H\alpha, int}$ and $A_{\rm H\alpha}$ as free parameters for individual spaxels. Maximizing $\ln\mathcal{P}_{\rm spaxel}$ for each spaxel will obtain the posterior probability of the ($\Delta m_{\rm X}$, $c$) pair, which is then summed up to give $\ln\mathcal{P}_{\rm sample}$. Finally, we maximize $\ln\mathcal{P}_{\rm sample}$ to obtain the best-fit ($\Delta m_{\rm X}$, $c$) pair for the whole sample. The \texttt{minimize} function from the \texttt{scipy} module is utilized to maximize $\ln\mathcal{P}_{\rm spaxel}$ for each spaxel. When maximizing $\ln\mathcal{P}_{\rm sample}$, we first use the \texttt{minimize} function to minimize $-\ln\mathcal{P}_{\rm sample}$ and obtain an initial guess of the best-fit parameters, which is then fed to the \texttt{emcee} module \citep{ForemanMackey2013} to perform a Markov chain Monte Carlo (MCMC; e.g., \citealt{Sharma2017}) sampling after adding small random perturbations. The medians and 16th-84th percentile ranges of the posterior probability distributions constructed from the MCMC chains after discarding a certain fraction of initial steps (i.e., the so-called ``burn-in'' phase) are taken as the final best-fit parameters and their uncertainties, respectively.

It is noteworthy that although we adopt a specific curve (i.e., the \citealt{Fitzpatrick1999} curve with $R_V=3.1$) as a zero point for the parameter, $\Delta m_{\rm X}$, during the fitting (see Equation \ref{eq: def_dm}), the best-fit slopes ($m_{\rm X}$) are not dependent on that choice. More specifically, although the definition of the relative slope $m_{\rm X}$ (Equation \ref{eq: def_m}) involves the attenuation difference between \ha\ and \hb, which is computed by assuming a certain attenuation curve, replacing this curve with other attenuation curves (e.g., the \citealt{Calzetti2000} curve) in the fitting would change the best-fit $\Delta m_{\rm X}$ but keep the best-fit $m_{\rm X}$ unchanged. The proof is given in Appendix \ref{appendix: proof_unchanged_mx}.

Our method has the following advantages. First, using the emission-line fluxes directly in the fitting instead of the logarithmic line ratios naturally avoids bias from the propagated uncertainties of line ratios which are asymmetric and non-Gaussian. Second, although the Case B recombination is assumed in the fitting, any deviations from it can be accounted for by the constant $c$ and thus do not influence the determination of $\Delta m_{\rm X}$. Other systematic uncertainties due to data reduction also can be handled in the same manner, as long as they are multiplicative in nature. Third, the inconsistency between the simulated median curves and their true values due to measurement uncertainties described in Section \ref{subsec: simu_geometry} can be naturally taken into account without the need to set a fitting range subjectively (see Section \ref{subsec: res_min-mcmc}).

\subsection{Simulation Validation for the Fitting Procedure}
\label{subsec: simu_fitting}

\subsubsection{Overall Performance at Different SNR Thresholds}
\label{subsubsec: simu_fitting_diffsnrcut}

\begin{figure*}[htb!]
    \centering
    \includegraphics[width=0.48\textwidth]{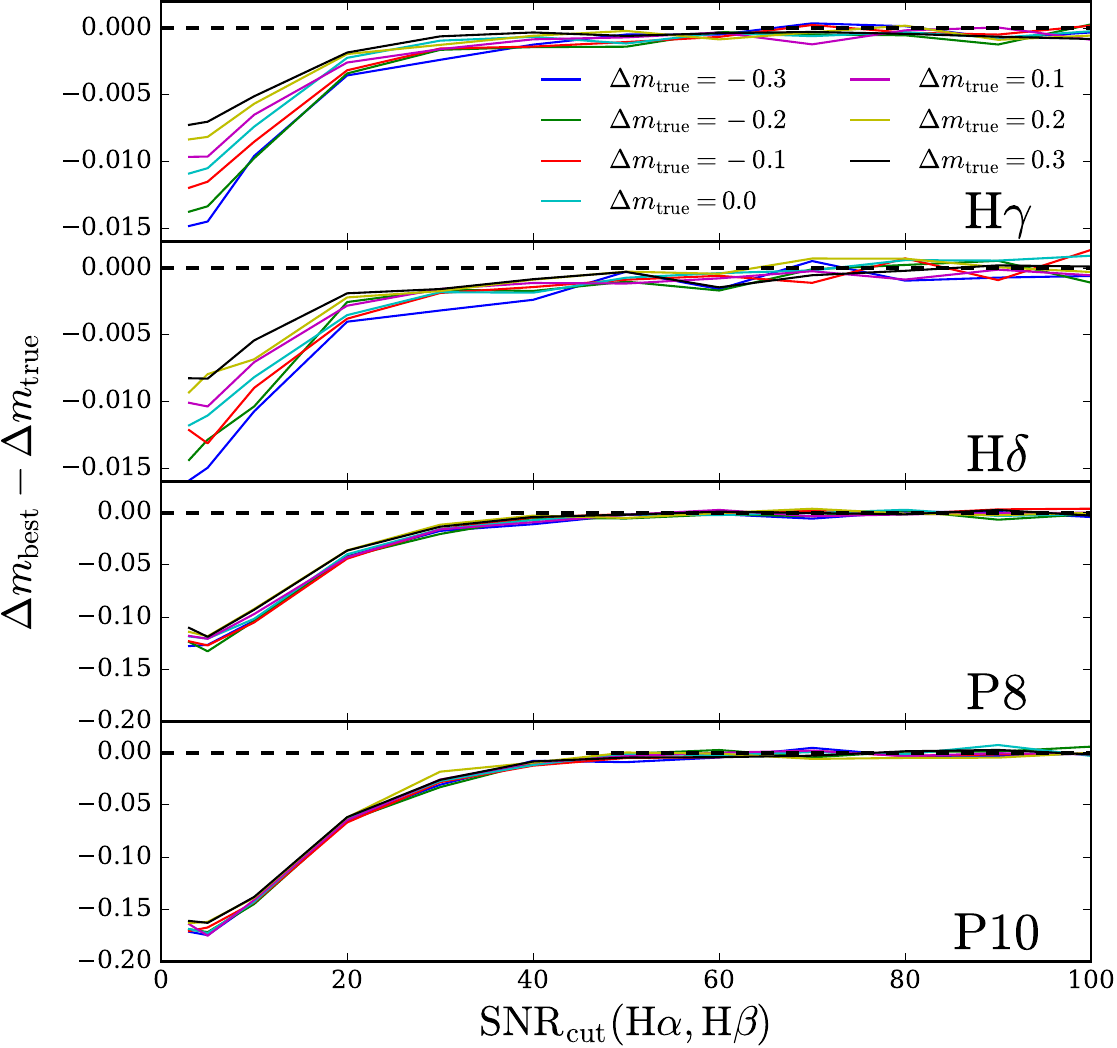}
    \includegraphics[width=0.48\textwidth]{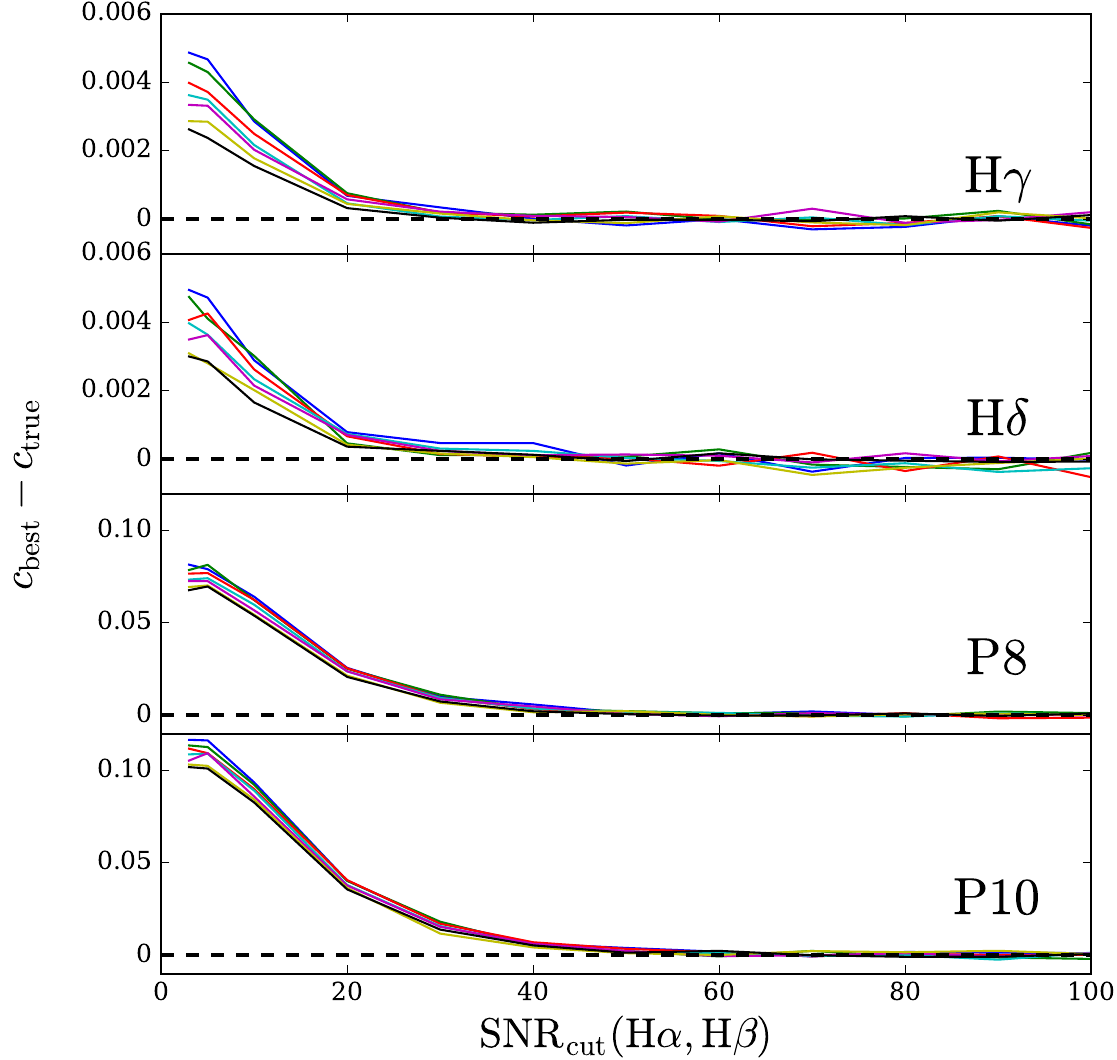}
    \caption{Differences between the recovered and true values of the free parameters $\Delta m$ (left) and $c$ (right) as a function of the adopted \ha\ and \hb\ SNR thresholds ranging from 3 to 100. Results based on the simulated catalogs with different true $\Delta m$ are delineated by different colors. The fitting procedure to derive the best-fit parameters only includes the \texttt{minimize} step for saving computing resources.
    \label{fig: sim_snrcut_dmc}}
\end{figure*}

To evaluate the recovery ability and possible systematic uncertainty of the fitting method, we generate simulated samples with different SNR thresholds for \ha\ and \hb, ranging from 3 to 100. The simulation procedure is the same as the one described in Section \ref{subsec: simu_geometry}, but only the foreground screen dust model is applied. To mock different extents of deviations from the Milky Way extinction curve, we adopt a series of true $\Delta m_{\rm X}$ ranging from -0.3 to 0.3 with a step of 0.1 for all targeted line X. No deviation from the Case B recombination is assumed, i.e., the true value $c_{\rm true}=1$. The simulated catalogs are then fed to the fitting process to obtain the best-fit parameter pair ($\Delta m_{\rm X}$, $c$) for each case. Because implementing the full \texttt{minimize}-MCMC fitting procedure described in Section \ref{subsec: fitting_method} is very expensive computationally, we only use the \texttt{minimize} step to derive the best-fit results of the simulated catalogs to save the computing time.

Fig. \ref{fig: sim_snrcut_dmc} exhibits how the differences between the recovered and true values of the parameters ($\Delta m_{\rm X}$ and $c$) vary as a function of the SNR cut applied to \ha\ and \hb. Note that for the targeted line X (i.e., \hg, \hd, P8, and P10), we do not apply any SNR cut. It is not a surprise to see that the deviations from the true values of the two parameters become significantly larger for simulated samples with lower SNR cut on \ha\ and \hb, regardless of the targeted lines. It seems that the fitting process provides an almost unbiased estimation of ($\Delta m_{\rm X}$, $c$) only for the simulated samples with $\mathrm{SNR_{cut}}\gtrsim 50$. Further inspection reveals that such deviations show very strong correlations with the fractions of spaxels that are set to zero fluxes in the simulated catalogs (see Section \ref{subsec: simu_geometry}). As the adopted SNR threshold decreases, more spaxels with weak recombination lines are included in the sample. As a result, the fractions of spaxels that have zero fluxes in the corresponding simulated catalogs increase and thus bias the distribution of the targeted line X, leading to a biased best-fit result. Since the zero-flux setting in the simulation aims to mock the results of the non-negative Gaussian fitting implemented in the DAP, similar situations should also exist in the observational data. Fig. \ref{fig: sim_snrcut_dmc} also implies that the deviations of both free parameters are dependent on the true $\Delta m_{\rm X}$ values, especially when the deviations are not too large (i.e., \hg\ and \hd), suggesting that applying a simple correction to the best-fit values to obtain an unbiased estimation seems to be infeasible. The coincident trends of the differences in $\Delta m_{\rm X}$ and $c$ reflect that the two parameters are degenerate to some extent, which can be confirmed via the posterior distributions (e.g., Fig. \ref{fig: res_minmc_bestfit}).

\subsubsection{Fitting Subsample Selected by SFR Surface Density}
\label{subsubsec: simu_fitting_fitsubsample}

As demonstrated in the last subsection, determining the nebular dust attenuation curve using the whole sample (i.e., star-forming spaxel sample with SNR threshold of 3 for \ha\ and \hb) would inevitably introduce substantial systematic bias that is difficult to correct. We thus consider constructing a more conservative subsample to perform the fitting. Simply adopting an SNR cut at 50 will leave only 9.5\% of the original sample, for which the sample size is too small and might not be representative of star-forming regions in the low-redshift Universe. On the other hand, since SNRs of emission lines strongly depend on the exposure time and observation conditions, we instead perform the subsample selection using another physical parameter, the deprojected SFR surface density ($\Sigma_{\rm SFR, cor}$). When computing $\Sigma_{\rm SFR, cor}$, the \cite{Fitzpatrick1999} Milky Way extinction curve with $R_V=3.1$ is adopted to correct dust attenuation of \ha\ fluxes. SFR is calculated from the dust-corrected \ha\ luminosity adopting the \cite{Kennicutt1998} conversion relation under the \cite{Salpeter1955} initial mass function assumption. We also retrieve the axis ratio from the NASA-Sloan Atlas catalog \citep{Blanton2011, Albareti2017}, which is derived from a single-component S\'{e}rsic fit in r-band, to correct for the projection effect due to the inclination (e.g., \citealt{Barrera-Ballesteros2016}).

\begin{figure}[htb!]
    \centering
    \includegraphics[width=0.48\textwidth]{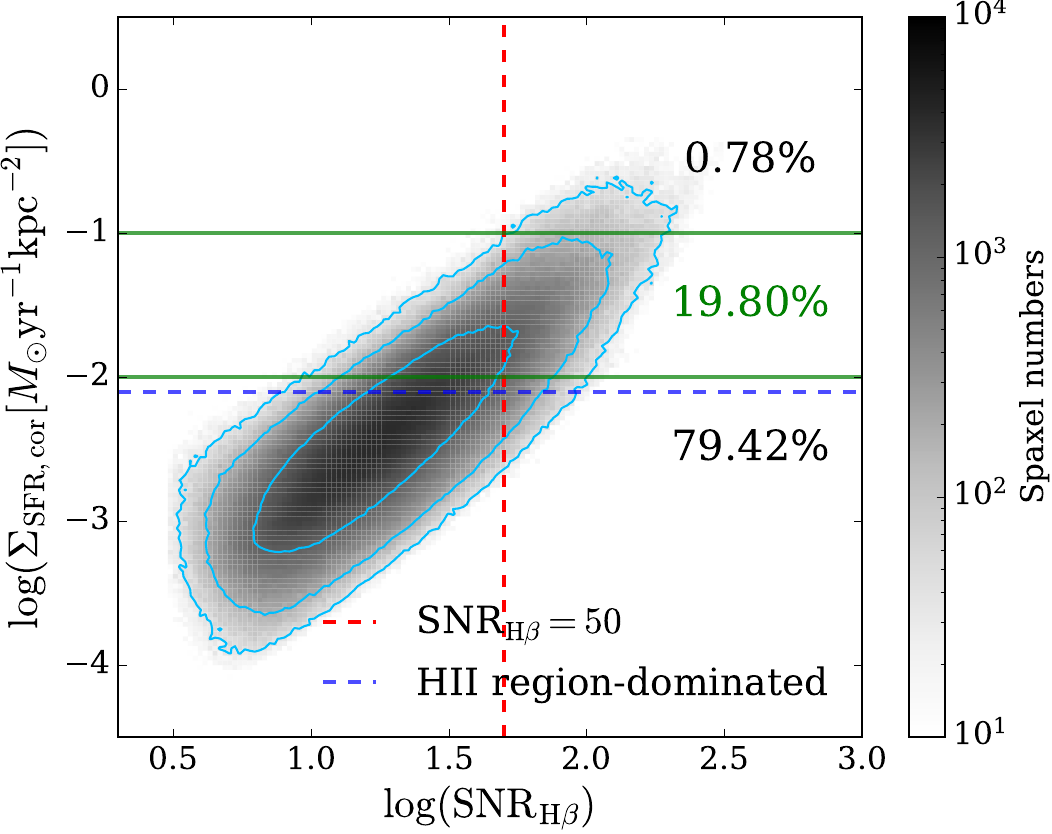}
    \caption{SNR of \hb\ as a function of the corrected SFR surface density. The vertical red dashed line indicates $\mathrm{SNR_{H\beta}}=50$, while the blue dashed line is the selection criterion of \hii\ region-dominated spaxels for the MaNGA survey proposed by \cite{Zhang2017}. The two green horizontal lines locate $\Sigma_{\rm SFR, cor}=0.01$ and 0.1 $M_{\odot}\mathrm{yr^{-1}kpc^{-2}}$, respectively. The three percentages suggest the fractions of spaxels within each $\Sigma_{\rm SFR, cor}$ bin. Spaxels between the green lines are adopted as the subsample to fit the slopes of the nebular dust attenuation curve. The underlying gray map and cyan contours are the same as those in Fig. \ref{fig: HII_diag}.
    \label{fig: SFRSDselection}}
\end{figure}

Although the SNR selection mentioned in Fig. \ref{fig: sim_snrcut_dmc} is applied to both the \ha\ and \hb\ lines, it should be dominated by the \hb\ threshold due to the weaker nature of \hb. We thus present the distribution of the main sample ($\mathrm{SNR_{cut}=3}$) on the SNR(\hb)--$\Sigma_{\rm SFR, cor}$ plane in Fig. \ref{fig: SFRSDselection} to illustrate our $\Sigma_{\rm SFR, cor}$-based sample selection. Our fitting subsample only includes $\sim 6.9\times 10^5$ spaxels with $-2\leq \log\Sigma_{\rm SFR, cor}[M_{\odot}~\mathrm{yr^{-1}~kpc^{-2}}]<-1$, which account for nearly 20\% of the main sample. Less than 1\% spaxels at the highest $\Sigma_{\rm SFR, cor}$ end are excluded due to the consideration of removing extreme regions. The lower limit of our $\Sigma_{\rm SFR, cor}$ criterion is very close to the suggested boundary between \hii-dominated and DIG-dominated spaxels for the MaNGA survey (blue dashed line in Fig. \ref{fig: SFRSDselection}) proposed by \cite{Zhang2017}, making sure that spaxels in this subsample are mainly excited by young massive OB stars. Although only 38.7\% of spaxels in this $\Sigma_{\rm SFR, cor}$-selected sample have SNR(\hb) $> 50$, we will demonstrate in the following that a nearly unbiased estimation of $\Delta m_{\rm X}$ can be derived from this subsample.

\begin{figure*}[htb!]
    \centering
    \includegraphics[width=\textwidth]{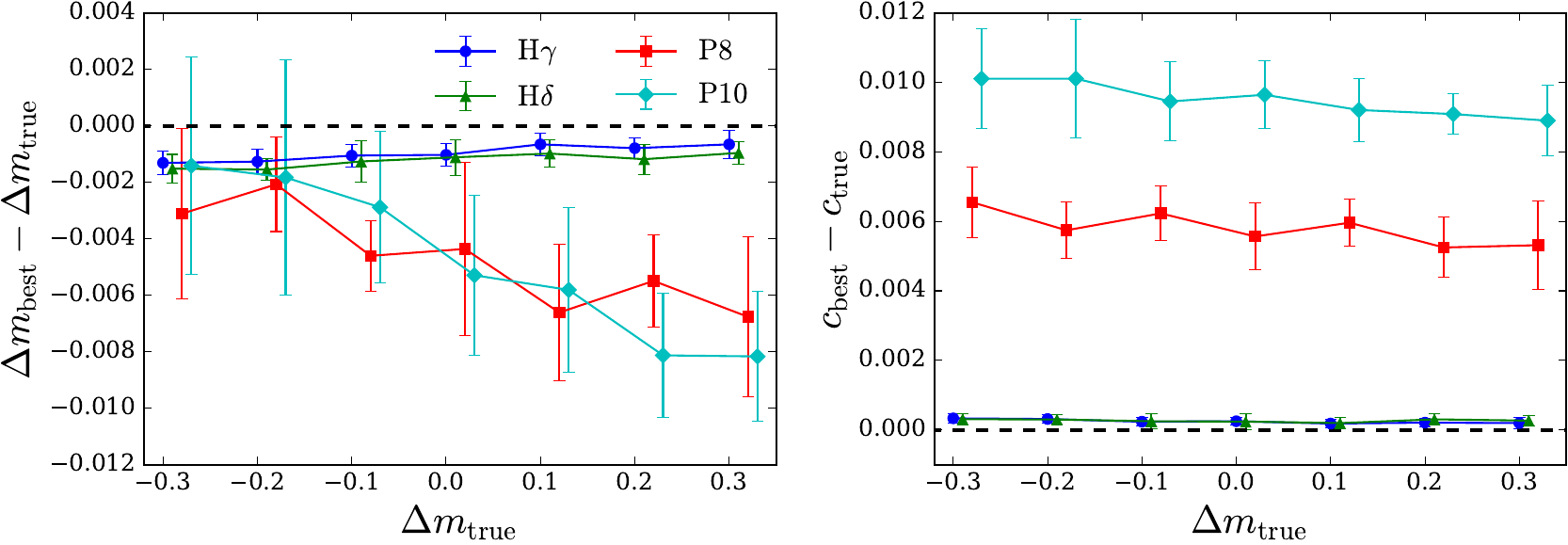}
    \caption{Differences between the true and recovered $\Delta m$ (using the \texttt{minimize} step only) as a function of the true $\Delta m$ for our $\Sigma_{\rm SFR, cor}$-selected sample. The symbols with error bars indicate the means and standard deviations of the best-fit parameters for the 10 simulated samples in each case, respectively.
    \label{fig: sim_dmc_SFRSDsel}}
\end{figure*}

Based on the new fitting subsample, we generate 10 simulated catalogs following the procedure described in Section \ref{subsec: simu_geometry} but with different random seeds. Again, the true $\Delta m_{\rm X}$ varies from -0.3 to 0.3 with a step of 0.1 and $c_{\rm true}=1$. The means and standard deviations of the differences between the true values and the best-fit ones derived from the minimization step for the $\Sigma_{\rm SFR, cor}$-selected sample are presented in Fig. \ref{fig: sim_dmc_SFRSDsel}.

Compared with the true $\Delta m_{\rm X}$ values, we still obtain best-fit results with deviations for the fitting sample. However, such deviations are greatly reduced by more than one order of magnitude compared to the simulated results for the $\mathrm{SNR_{cut}}=3$ samples presented in Fig. \ref{fig: sim_snrcut_dmc} and will be treated as corrections to our best-fit results derived from the minimization-MCMC procedure. Even for the Paschen lines, the deviations in the relative slopes are $\lesssim 0.008$ in all cases, while the deviations in the scaling factor $c$ are smaller than 1\%.

From the posterior distributions provided by the MCMC, we can calculate the statistical uncertainties of the best-fit parameters, which is due to the emission-line measurement uncertainties and the size of the fitting sample. On the other hand, systematic uncertainties arising from imperfect data analysis, the adopted assumptions, and/or the fitting method cannot be assessed from the MCMC procedure and are difficult to account for. However, given the simulation described here, we can quantify these uncertainties to some extent. To this end, we compute the average deviations in $\Delta m_{\rm X}$ for each targeted line by averaging over all assumed true $\Delta m_{\rm X}$. Note that the subtle trends of larger true $\Delta m_{\rm X}$ values having larger deviations for P8 and P10 are ignored due to the small variations over the simulated $\Delta m_{\rm X}$ range. The means and standard deviations of the deviations in $\Delta m_{\rm X}$ are $-0.001\pm0.001$, $-0.001\pm0.001$, $-0.005\pm0.003$, and $-0.005\pm0.004$ for \hg, \hd, P8, and P10, respectively. The standard deviations for \hg\ and \hd\ are smaller than 0.0005 but are rounded to 0.001 for conservative estimation of the uncertainties. Other hydrogen recombination lines (e.g., from H7 to H12 and P9) are also considered in the simulation and processed in the same manner, and the corresponding average deviations from the true values are given in Table \ref{tab: bestfit_paras}.

\subsection{Minimization-MCMC Results for the $\Sigma_{\rm SFR, cor}$-selected Sample}
\label{subsec: res_min-mcmc}

\begin{figure*}[htb!]
    \centering
    \includegraphics[width=0.49\textwidth]{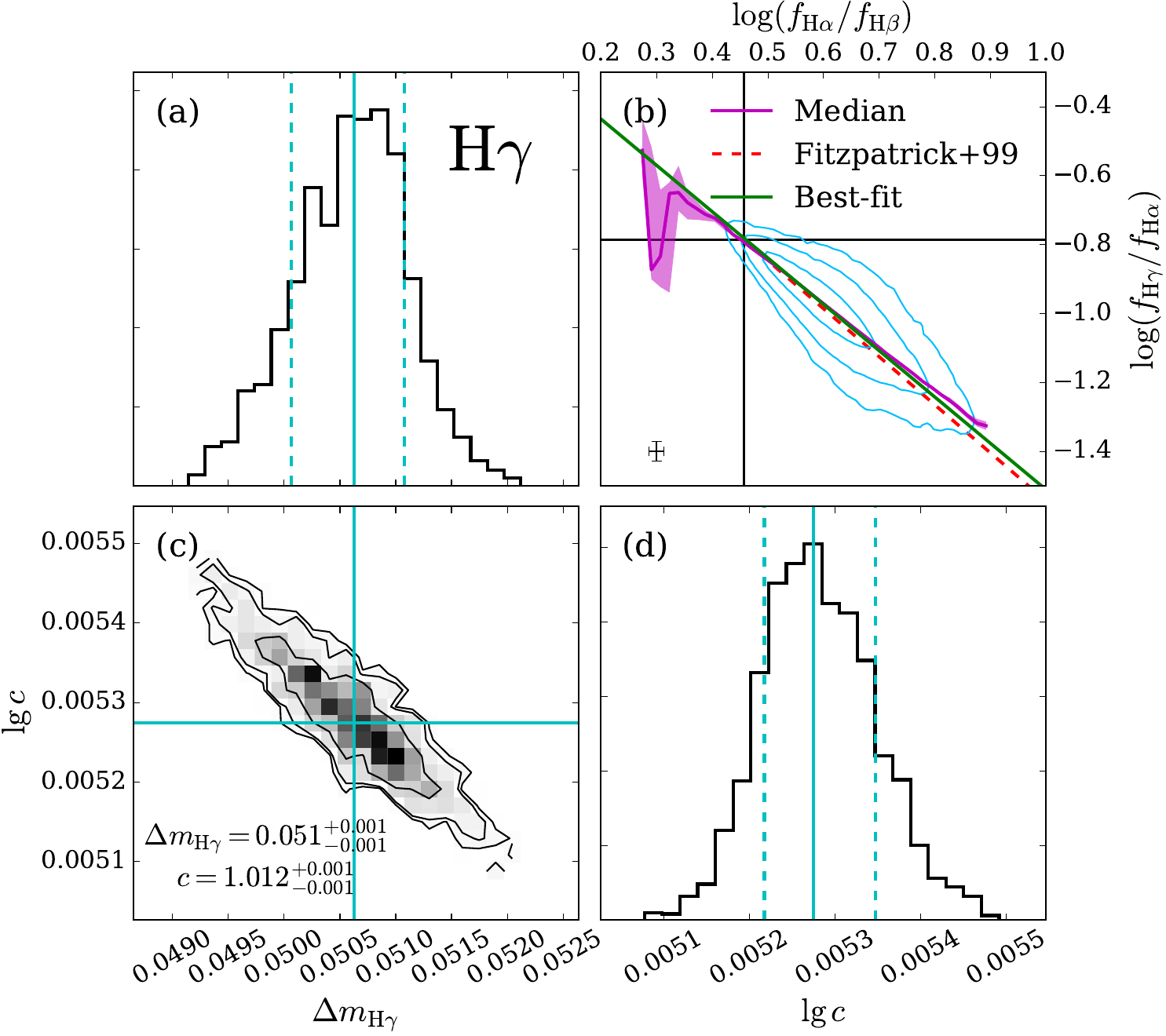}
    \includegraphics[width=0.49\textwidth]{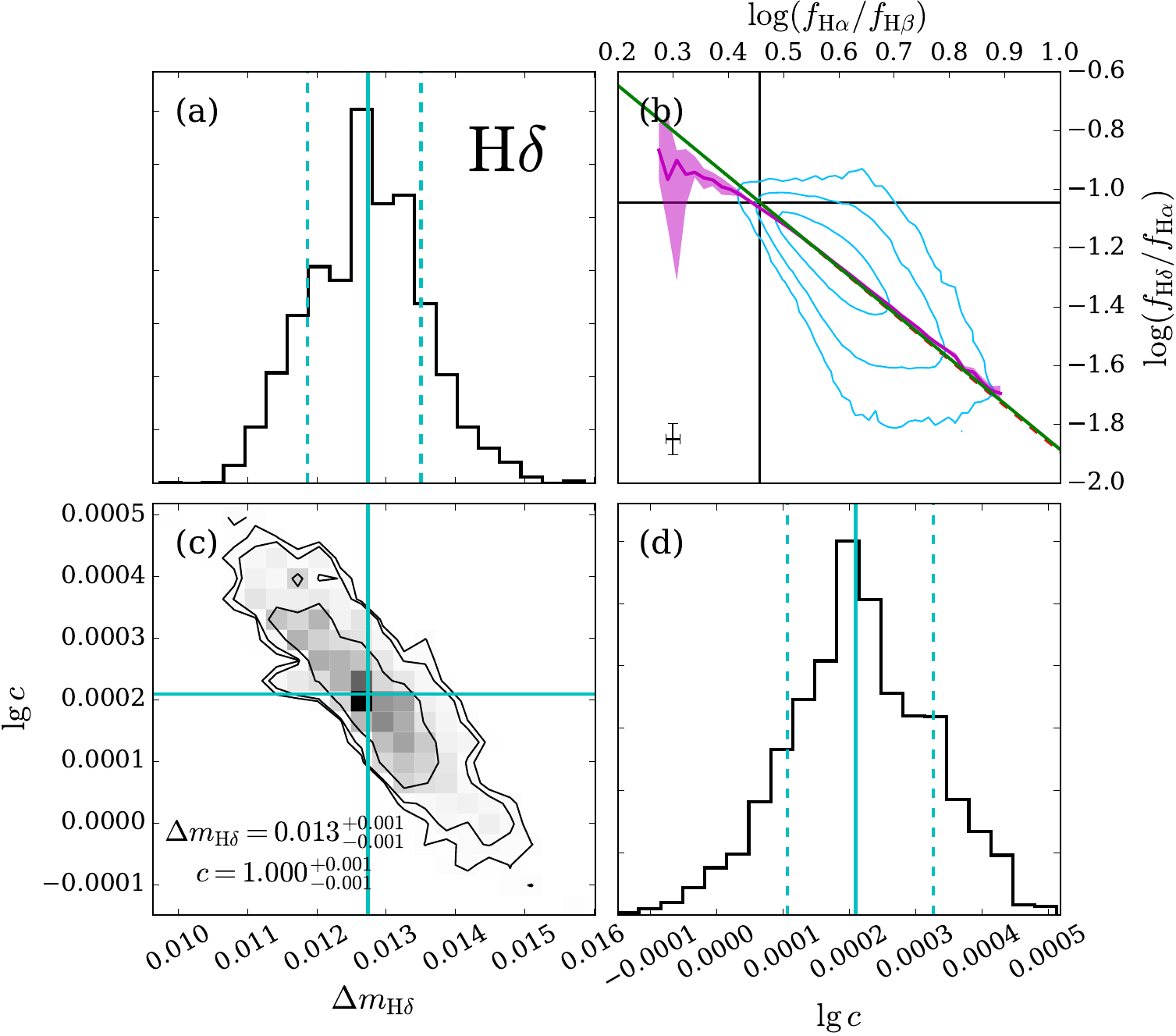}
    \includegraphics[width=0.49\textwidth]{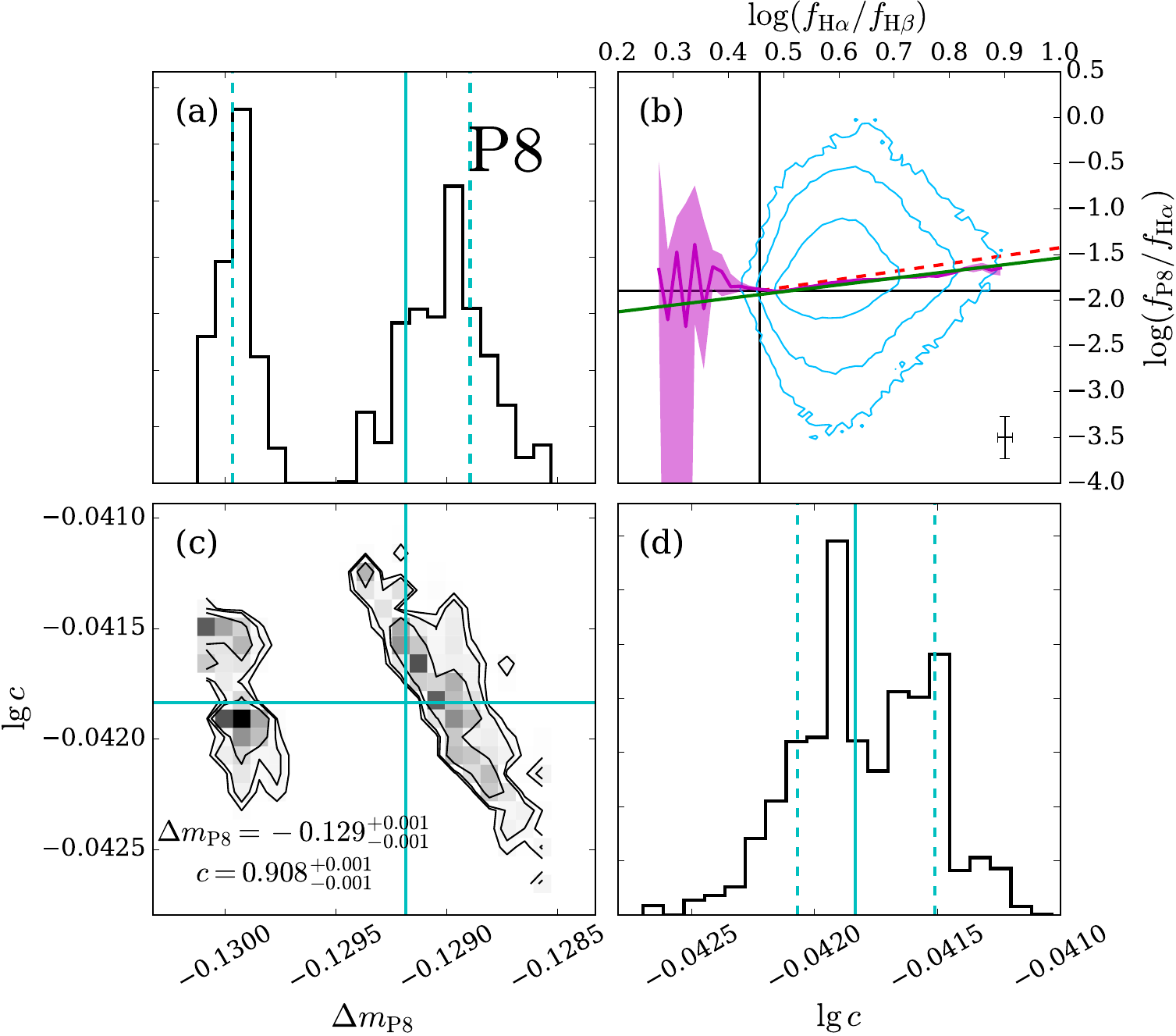}
    \includegraphics[width=0.49\textwidth]{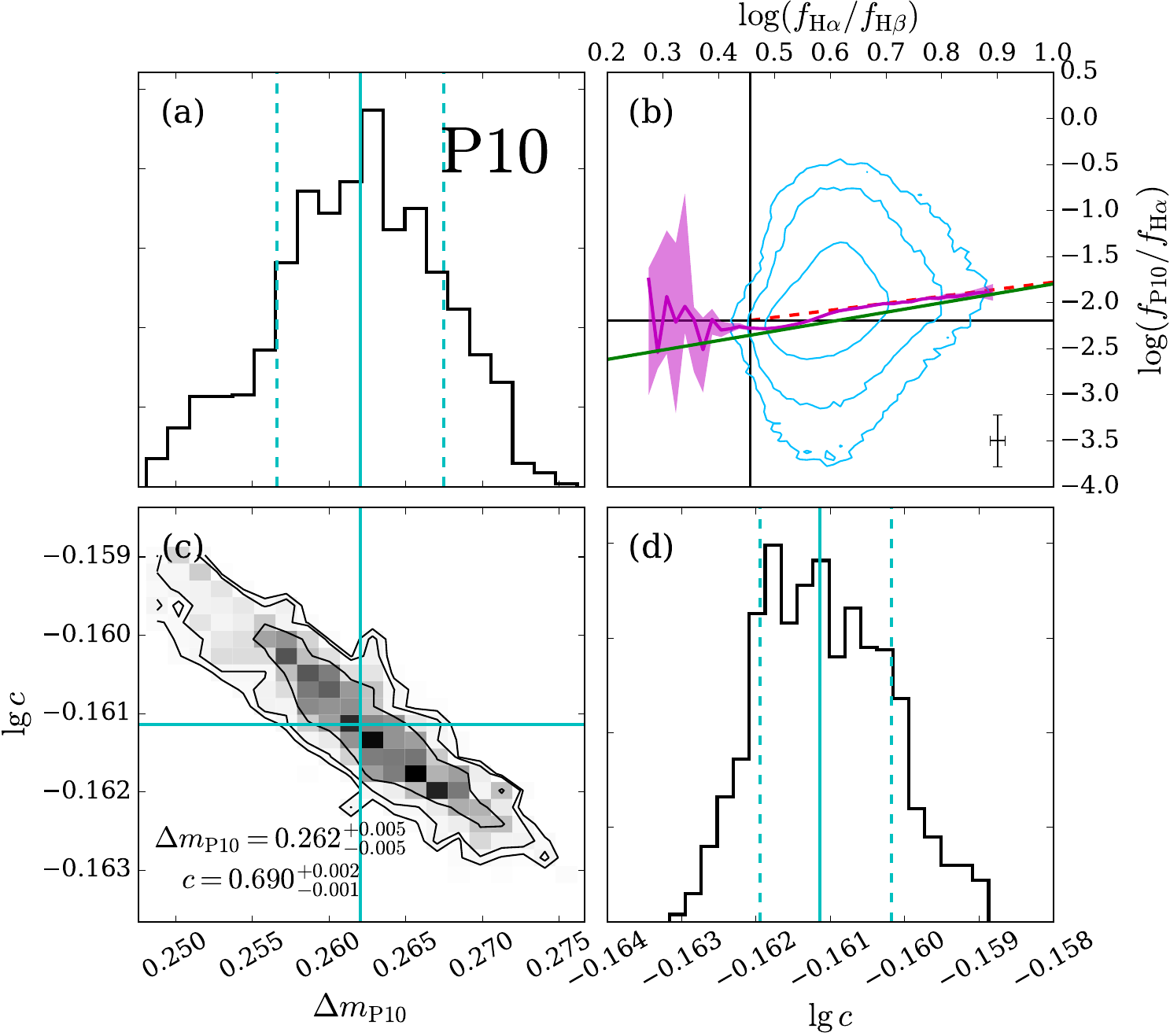}
    \caption{Best-fit results for the four main recombination lines derived from the minimization-MCMC fitting method. In each case, Panel (c) shows the posterior probability distribution of the two parameters, while Panels (a) and (d) are the marginalized distributions of individual parameters. The cyan solid lines in these panels locate the medians of the corresponding distributions, and the cyan dashed lines enclose the 16th-84th percentile ranges of the distributions. The fit-best results with 1$\sigma$ uncertainties (i.e., the 16th-84th percentile ranges) for $\Delta m_{\rm X}$ and $c$ are listed in Panel (c). Panel (b) gives the comparison between the best-fit relation (green line) and the observed median curve (magenta curve) on the line ratio plane. Other symbols are similar to those in Fig. \ref{fig: lineratio} but for the $\Sigma_{\rm SFR, cor}$-selected sample.
    \label{fig: res_minmc_bestfit}}
\end{figure*}

\begin{table*}
   \caption{Best-fit parameters from the minimization-MCMC procedure}
   \label{tab: bestfit_paras}
   \centering
   \begin{tabular}{cccccccc}
   \hline\hline
   Lines & $\Delta m_{\rm MCMC}$ & $\Delta m_{\rm best}-\Delta m_{\rm true}$ & $\Delta m_{\rm final}$ & $c_{\rm MCMC}$ & $c_{\rm best}-c_{\rm true}$ & $c_{\rm final}$ & $m_{\rm final}$ \\
    \hline
    \hg & $0.051\pm0.001$ & $-0.001\pm0.001$ & $0.052\pm0.001$ & $1.012\pm0.001$ & $0.000\pm0.001$ & $1.012\pm0.001$ & $-1.344$ \\
    \hd & $0.013\pm0.001$ & $-0.001\pm0.001$ & $0.014\pm0.001$ & $1.000\pm0.001$ & $0.000\pm0.001$ & $1.000\pm0.001$ & $-1.549$ \\
    P8 & $-0.129\pm0.001$ & $-0.005\pm0.003$ & $-0.124\pm0.003$ & $0.908\pm0.001$ & $0.006\pm0.001$ & $0.902\pm0.001$ & $0.743$ \\
    P10 & $0.262\pm0.005$ & $-0.005\pm0.004$ & $0.267\pm0.007$ & $0.690\pm0.002$ & $0.009\pm0.001$ & $0.681\pm0.002$ & $1.029$ \\
    H7 & $0.016\pm0.001$ & $-0.001\pm0.001$ & $0.017\pm0.002$ & $0.997\pm0.001$ & $0.000\pm0.001$ & $0.997\pm0.001$ & $-1.639$ \\
    H8 & $0.614\pm0.014$ & $0.180\pm0.031$ & $0.434\pm0.034$ & $0.534\pm0.002$ & $-0.003\pm0.004$ & $0.537\pm0.005$ & $-1.280$ \\
    H9 & $0.196\pm0.004$ & $0.013\pm0.004$ & $0.183\pm0.006$ & $0.840\pm0.001$ & $-0.002\pm0.001$ & $0.842\pm0.001$ & $-1.570$ \\
    H10 & $0.159\pm0.007$ & $0.043\pm0.010$ & $0.116\pm0.012$ & $0.652\pm0.001$ & $-0.005\pm0.002$ & $0.657\pm0.002$ & $-1.665$ \\
    H11 & $1.034\pm0.008$ & $0.096\pm0.021$ & $0.938\pm0.022$ & $0.552\pm0.002$ & $-0.008\pm0.003$ & $0.561\pm0.003$ & $-0.864$ \\
    H12 & $1.342\pm0.010$ & $0.165\pm0.033$ & $1.177\pm0.034$ & $0.592\pm0.002$ & $-0.009\pm0.005$ & $0.601\pm0.005$ & $-0.641$ \\
    P9 & $-0.254\pm0.005$ & $-0.006\pm0.004$ & $-0.248\pm0.007$ & $1.017\pm0.002$ & $0.009\pm0.001$ & $1.008\pm0.002$ & $0.559$ \\
    \hline
   \end{tabular}
   \tablefoot{For each parameter ($\Delta m_{\rm X}$ and $c$), the three columns from left to right are the best-fit results from the MCMC step, the differences between the best-fit results and the true values computed from the simulation (i.e., systematic uncertainties), and the final estimations, respectively. The final column is the final estimation of $m_{\rm X}$ after adding the \cite{Fitzpatrick1999} value of each recombination line. We do not list the uncertainties of $m_{\rm final}$ since they are the same as those of $\Delta m_{\rm final}$. Although the \cite{Fitzpatrick1999} extinction curve is used as the reference here, the results on $m_{\rm final}$ are independent of the choice of the reference curve.}
\end{table*}

We apply the minimization-MCMC fitting procedure to the $\Sigma_{\rm SFR, cor}$-selected sample to determine the best-fit relative slopes for star-forming regions on kpc scales. The best-fit results for the four main targeted lines are given in Fig. \ref{fig: res_minmc_bestfit}. For each targeted line, we provide a corner plot to display the posterior distributions of the parameters and an \ha/\hb--X/\ha\ line ratio diagram to show the comparison between the observed median curve and the prediction based on the best-fit parameters. From the fitting, we take the medians and the 16th-84th percentile ranges of the posterior distributions as the best-fit parameters and 1$\sigma$ uncertainties, respectively, for $\Delta m_{\rm X}$ and $c$.

For \hg\ and \hd, the scaling factors are $\lesssim 1\%$, which are much smaller than the median measurement uncertainties of these two lines in the fitting subsample. These negligible deviations for predicted fluxes demonstrate the reliability of the Case B recombination assumption for most of the spaxels in our sample and the accuracy of MaNGA's relative flux calibration among these lines.\footnote{It might be possible that both cases are not true and bias the predicted fluxes by a similar amount but with different signs, and thus a $\sim 1$ scaling factor is still returned by the fitting procedure. However, it is nearly impossible that this coincidence happens simultaneously for three emission lines (i.e., \hg, \hd, and H7, see Table \ref{tab: bestfit_paras} and Fig. \ref{fig: lineratio_bestfit_others})}
Hence, the relative slopes for these two lines can be well constrained. However, the best-fit $c$ values for P8 and P10 suggest that the predicted fluxes from the Case B recombination should be reduced by about 9\% and 31\%, respectively, to match the observations. Thus, we have more complex situations for these two lines. We will discuss possible reasons for these systematic biases in Section \ref{subsec: reasons_c_deviations}. Intriguingly, the posterior distribution of the two parameters for P8 has a double-peak feature, leading to bimodal probability distributions for individual parameters. Although we have no idea about the reason for this odd feature that is different from other emission lines (also see Fig. \ref{fig: lineratio_bestfit_others}), the overall parameter space is still narrow and results in 1$\sigma$ uncertainties of $<0.001$ for both parameters.

The best-fit $\Delta m_{\rm X}$ are very small for \hg\ and \hd, while larger deviations are obtained for two Paschen lines. Although we do not fit the median curves directly, the best-fit relations between the line ratios are generally consistent with the median ones at $f_{\rm H\alpha}/f_{\rm H\beta}\gtrsim 2.86$ except for the P10 line. If the best-fit relations for \hg, \hd, and P8 are treated as the ``true'' values of the fitting sample, the offsets of the observed median curves from these ``true'' values are qualitatively in line with the ones between the simulated median curves and the foreground screen model predictions (i.e., also the true values in the simulation) shown in Fig. \ref{fig: lineratio_sim}. This suggests that the scattering in line ratios due to the measurement uncertainties of emission-line fluxes discussed in Section \ref{subsec: simu_geometry} is naturally taken into account by our fitting approach. As a result, an unbiased estimate of the slope can be obtained. This is one of our motivations for developing a new fitting method.

As mentioned above, the uncertainties obtained from this procedure are only the statistical ones accounting for the uncertainties from emission-line measurements. As shown in Fig. \ref{fig: res_minmc_bestfit}, the derived uncertainties from this source for both free parameters are very small, regardless of the targeted lines, due to the large number of spaxels in our fitting sample. Together with the corrections and systematic uncertainties derived from Section \ref{subsubsec: simu_fitting_fitsubsample}, we can obtain the final estimations for $\Delta m_{\rm X}$ and $c$ and list them in Table \ref{tab: bestfit_paras}. The best-fit parameters from the minimization-MCMC procedure only, the differences between the best-fit and true values based on the simulation (i.e., systematic uncertainties), and the final estimation of the relative slope ($m_{\rm final}$) for each recombination line are also given. We emphasize that, theoretically, the derived $m_{\rm final}$ is independent of the reference extinction/attenuation curve adopted in the fitting (see Appendix \ref{appendix: proof_unchanged_mx}) and thus can be directly compared with the predictions of different curves. For simplicity, the asymmetric uncertainties drawn from the posterior distributions in the MCMC fitting step are adjusted to be symmetric by adopting the larger side, although the exact asymmetric uncertainties are still listed in Fig. \ref{fig: res_minmc_bestfit}.

We find that the deviation of $m_{\rm H\gamma}$ from the \cite{Fitzpatrick1999} value is small but significant given the very small uncertainty. When applying to a region with $E(B-V)=0.6$ mag,\footnote{Note that the $E(B-V)$ of 99\% spaxels in the $\mathrm{SNR_{cut}}=3$ sample are $<0.6$ mag when a \cite{Fitzpatrick1999} extinction curve is assumed.} this could result in a difference of about 3.6\% in the dust correction factor compared to the standard \cite{Fitzpatrick1999} extinction curve with $R_V=3.1$.
The final best-fit $\Delta m_{\rm H\delta}$ ($\Delta m_{\rm H7}$) only contributes to an additional 1.0\% (1.2\%) of the dust correction in the same region. Such small corrections (together with the result for H7, see Section \ref{appendix: res_otherlines}) imply that the nebular dust attenuation on kpc scales can be described by the \cite{Fitzpatrick1999} extinction curve within an accuracy of about 4\% in terms of the dust correction factor. However, when comparing with the \cite{Calzetti2000} attenuation curve with $R_V=4.05$, similar corrections in the dust corrected fluxes are 5.1\%, 5.6\%, and 8.1\% for \hg, \hd, and H7, respectively. For the Paschen lines, due to the degeneracy between $\Delta m_{\rm X}$ and $c$ and the large departures of $c$ from 1, their slopes could not be considered as well constrained before the sources of the non-unity $c$ are resolved and quantified. Therefore, we conclude that only $m_{\rm X}$ of strong recombination lines like \hg, \hd, and H7 can be well constrained given the current emission-line measurements.

\subsection{Possible Reasons for the Non-unity Scaling Factors}
\label{subsec: reasons_c_deviations}

Both Fig. \ref{fig: res_minmc_bestfit} and Table \ref{tab: bestfit_paras} show that a scaling factor $c$ smaller than 1 is required for P8 and P10 to let the predicted fluxes match with the observations. Possible reasons that might contribute to these systematic reductions are deviations from the Case B recombination assumption, improper stellar continuum subtraction, residuals of night-sky lines, improper emission-line measurements, and so on. Given the excellent agreements with Case B for \hg, \hd, and H7 (see Table \ref{tab: bestfit_paras}), it is unlikely that this large offset from 1 is due to the intrinsic deviations from Case B predictions.

Because the sky residuals at the red end (around the Paschen lines) are still observable in some spectra, we have tried to remove this contamination by constructing an average template of night-sky line residuals for each Paschen line using emission-line fluxes from the DAP catalog. The resulting templates exhibit abundant night-sky lines, most of which are much stronger than the flux levels of the Paschen lines extracted from the DAP catalog. Although the fitting using the line fluxes after correcting for these sky residuals can have slightly smaller reduced $\chi^2$, the large uncertainties of such corrections are difficult to estimate. We thus do not apply these corrections in the fitting results presented in this work and leave this issue for future study in which we will construct the templates from the MaNGA spectra directly (e.g., \citealt{Yan2011}).

Another important effect of the Paschen lines is the zero-flux problem. Since we do not apply SNR cut for our targeted lines, all spaxels with unmasked Gaussian fitting fluxes are included in the sample. For very weak emission lines, DAP might return zero fluxes with non-zero uncertainties. As a result, even for the $\Sigma_{\rm SFR, cor}$-selected subsample, the fractions of spaxels with zero fluxes are about 18\% and 15\% for P8 and P10, respectively.\footnote{The P10 line should have a smaller intrinsic flux but suffer more dust attenuation compared to the P8 line. Contrary to what is observed, we thus should expect more spaxels with zero fluxes for P10 rather than P8. This conflict might reflect the contamination of the sky residuals.}
If the emission-line fluxes are measured by the means of fitting by a non-negative single Gaussian profile, the zero-flux problem is inevitable and significant given the faintness of the Paschen lines. The problem would depress the observed fluxes systematically and thus lead to a scaling factor smaller than 1, which aligns with our fitting results shown in Fig. \ref{fig: res_minmc_bestfit}. However, the dominant effect is difficult to determine and resolve with the current data.

The best-fit results listed in Table \ref{tab: bestfit_paras} for other recombination lines also demonstrate that the non-unity scaling factor problem also exists for the high-order Balmer lines (except H7) and P9. The P9 line has a very similar situation to those of P8 and P10. For the high-order Balmer series, we defer the discussion to Appendix \ref{appendix: res_otherlines} and briefly summarize here that either the imperfect stellar continuum modeling or the zero-flux problem of weak lines might play an important role in the scaling factor $c$.

\section{Summary}
\label{sec: summary}

In this work, we use the complete data release of the MaNGA survey to investigate the nebular dust attenuation curve for star-forming regions on kpc scales. We examine the dust geometry models observed by the hydrogen recombination lines via simple simulations. We propose a novel approach to fit the relative slopes of the nebular dust attenuation curve with emission-line fluxes directly and apply it to a $\Sigma_{\rm SFR, cor}$-selected subsample. Our main results are as follows.
\begin{enumerate}\renewcommand\labelenumi{(\theenumi)}
    \item Most of the star-forming regions favor the foreground screen dust model on kpc scales rather than the uniform mixture model.
    \item Based on an ideal simulation considering Gaussian statistical flux uncertainties only, we find that our minimization-MCMC fitting procedure can recover the true values for both the slopes of the nebular dust attenuation curve and the scaling factors with nearly no bias for \hg, \hd, P8, and P10 for either high-SNR sample or the corresponding $\Sigma_{\rm SFR, cor}$-selected sample.
    \item By applying the method to the MaNGA data (i.e., the $\Sigma_{\rm SFR, cor}$-selected subsample), we find that the relative slope of the nebular attenuation curve can be well determined for strong hydrogen recombination lines (e.g., \hg, \hd, and H7). However, severe contaminations/systematic uncertainties prevent us from obtaining reasonable values of the slopes and intercepts for weak recombination lines (e.g., the high-order Balmer lines or the Paschen lines).
    \item For the \hii\ region-dominated spaxels, the relative slopes of the nebular attenuation curve at the wavelengths of \hg, \hd, and H7 can be well described by the \cite{Fitzpatrick1999} Milky Way extinction curve with $R_V=3.1$ with an accuracy in the corrected flux of $\lesssim 4\%$.
    \item Implied by both the simulation and fitting results, emission-line fluxes obtained from non-negative Gaussian fitting are biased in the weak-line regime.
\end{enumerate}

Our results further reveal that although the emission-line measurements from the DAP are very accurate for strong lines, the measurements for weak lines still suffer many problems, e.g., significant residuals of the night-sky line at the red end of spectra and improper stellar continuum subtraction at the blue end. The Gaussian fitting might also be problematic for weak lines where the fitting process might fit a nearby noise or directly return zero fluxes. To mitigate this problem, we are currently using optimal extraction (e.g., \citealt{Horne1986}) to produce unbiased flux measurements for emission lines in the MaNGA spectra and hope to have a better flux estimator for weak lines. Overall, more accurate emission-line measurements are needed to better constrain the shape of the nebular dust attenuation curve.

\begin{acknowledgements}
    We would like to thank the anonymous referee for the very helpful comments. Z.S.L. would like to thank Xihan Ji for his valuable discussion during the preparation of this work. We acknowledge the support by the Research Grant Council of Hong Kong (Project No. 14302522) and the National Natural Science Foundation of China (NSFC; grant No. 12373008). Z.S.L. acknowledges the support from Hong Kong Innovation and Technology Fund through the Research Talent Hub program (PiH/022/22GS). RY acknowledges support by the Hong Kong Global STEM Scholar Scheme (GSP028) and by the Hong Kong Jockey Club Charities Trust through the project, JC STEM Lab of Astronomical Instrumentation and Jockey Club Spectroscopy Survey System.\\

    Funding for the Sloan Digital Sky Survey IV has been provided by the Alfred P. Sloan Foundation, the U.S. Department of Energy Office of Science, and the Participating Institutions. SDSS-IV acknowledges support and resources from the Center for High Performance Computing at the University of Utah. The SDSS website is www.sdss4.org. SDSS-IV is managed by the Astrophysical Research Consortium for the Participating Institutions of the SDSS Collaboration including the Brazilian Participation Group, the Carnegie Institution for Science, Carnegie Mellon University, Center for Astrophysics | Harvard \& Smithsonian, the Chilean Participation Group, the French Participation Group, Instituto de Astrof\'isica de Canarias, The Johns Hopkins University, Kavli Institute for the Physics and Mathematics of the Universe (IPMU) / University of Tokyo, the Korean Participation Group, Lawrence Berkeley National Laboratory, Leibniz Institut f\"ur Astrophysik Potsdam (AIP),  Max-Planck-Institut f\"ur Astronomie (MPIA Heidelberg), Max-Planck-Institut f\"ur Astrophysik (MPA Garching), Max-Planck-Institut f\"ur Extraterrestrische Physik (MPE), National Astronomical Observatories of China, New Mexico State University, New York University, University of Notre Dame, Observat\'ario Nacional / MCTI, The Ohio State University, Pennsylvania State University, Shanghai Astronomical Observatory, United Kingdom Participation Group, Universidad Nacional Aut\'onoma de M\'exico, University of Arizona, University of Colorado Boulder, University of Oxford, University of Portsmouth, University of Utah, University of Virginia, University of Washington, University of Wisconsin, Vanderbilt University, and Yale University.
\end{acknowledgements}

\bibliographystyle{aa}

\begin{appendix}

    \section{Goodness of emission-line measurements}
    \label{appendix: other_effects}

    Due to the existence of stellar absorptions at the wavelengths of hydrogen recombination lines, the accuracy of flux measurements of these lines strongly depends on both the fitting to the underlying stellar continuum and the Gaussian fitting to the gas-only (i.e., continuum-subtracted) emission-line spectrum. \cite{Belfiore2019} provides very detailed assessments of the emission-line measurement in the MaNGA DAP and demonstrates statistically robust emission-line flux measurements for the vast majority of the MaNGA spectra. Here we provide further discussion about the goodness of the stellar-continuum subtraction.
    
    Based on idealized recovery simulations in which the same stellar templates were adopted to generate and fit the mock data, \cite{Belfiore2019} investigated how well the DAP can recover the input emission-line parameters. The authors concluded that the fluxes and the corresponding errors can have almost unbiased estimations down to S/N $\sim$ 1.5. However, when the ``wrong'' SSP templates are used in the fitting, they also showed that the measured fluxes could be systematically biased, highlighting the systematic errors arising from the stellar templates. A similar effect due to the adopted stellar templates was also reported by \cite{Groves2012a} where the EW(\hb) measurement was demonstrated to be biased and thus the Balmer decrement. Since we treat the \ha\ and \hb\ lines as a fiducial pair in our investigation, it is necessary to further examine the systematic effect due to the adopted SSP templates on the DAP flux measurements for these two lines.

    \begin{figure*}[htb!]
        \centering
        \includegraphics[width=\textwidth]{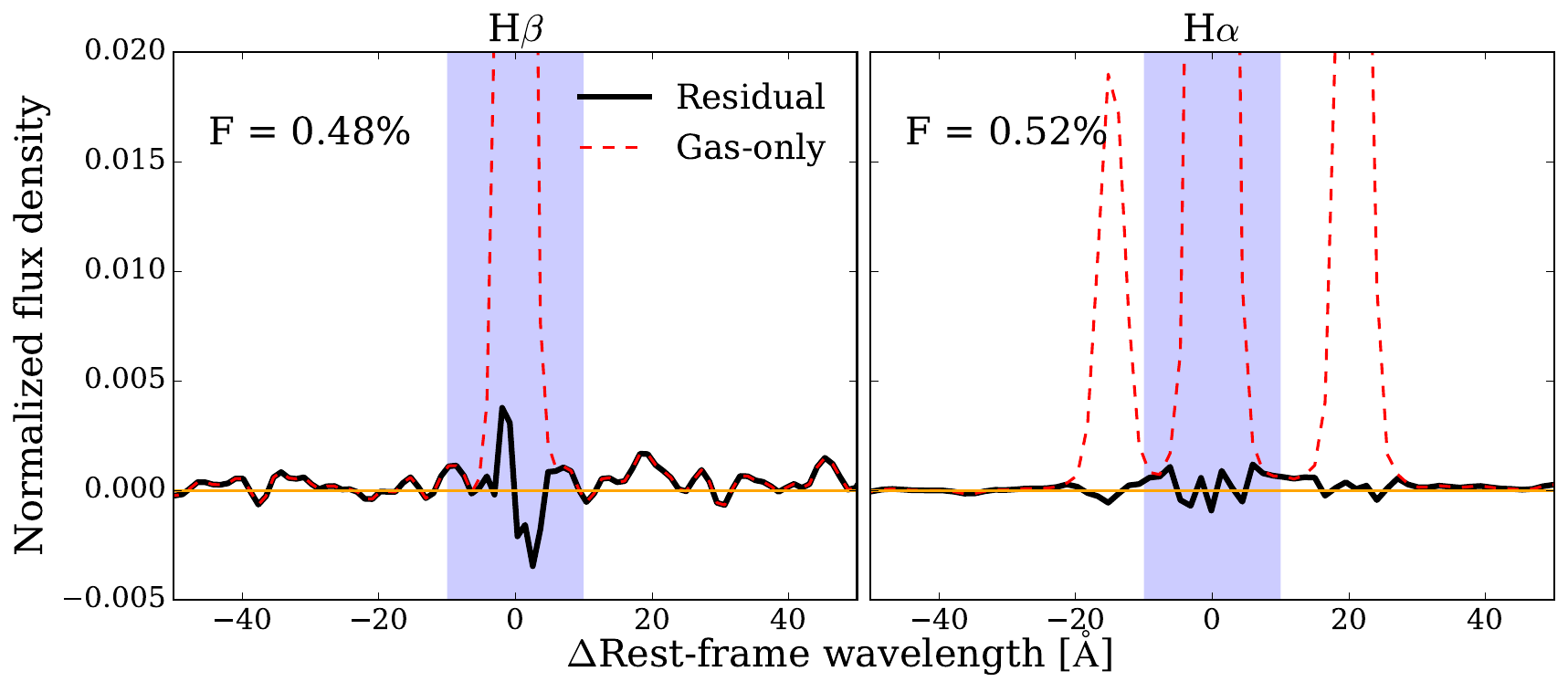}
        \caption{Stacked residual spectra normalized by emission-line flux for \hb\ (left) and \ha\ (right). The black solid curves are the normalized residual spectra, while the red dashed curves are the normalized gas-only (i.e., stellar-continuum subtracted) spectra for reference. The orange horizontal lines denote where the flux is zero. The F values (upper-left corners) indicate the fraction of integrated residual flux compared to the emission line flux, measured within the blue-shaded region of 20 \AA width.
        \label{fig: res_spec}}
    \end{figure*}

    Unfortunately, given that the ``true'' stellar templates of the observed spectra are unknown, direct comparison with the ones derived from the fitting involving ``true'' templates is impossible. Thanks to the complete DAP output, we here provide an indirect examination using the residual spectra after removing the best-fit models of both stellar continuum and emission lines. The aim of this calculation is to check whether there is a significant non-zero flux on the residual spectrum at the location of \ha\ or \hb\ line. We use the main sample (i.e., the SNR-selected star-forming sample) described in Section \ref{subsec: sample} and retrieve their residual spectra from the DAP output, which are then de-reshifted, resampled to the same wavelength grid, normalized by the measured fluxes of the lines we want to check (i.e., \ha\ or \hb). The median spectra of these \ha- or \hb-flux normalized residual spectra are exhibited in Fig. \ref{fig: res_spec}. We also provide the median stacked spectra of the gas-only spectra (i.e., only removing the best-fit stellar continuum models from the observed spectra) for reference in the figure. We calculate the integrated fluxes over a wavelength window of 20 \AA\ around the examined lines, which is large enough to enclose the lines (see the blue shaded region in Fig. \ref{fig: res_spec}), resulting in fractional residual fluxes of 0.48\% and 0.52\% for \hb\ and \ha, respectively. Such residual fluxes are much smaller than either the absolute or relative flux calibration of the MaNGA spectra \citep{Yan2016b} and thus can be ignored.
    
    On the other hand, to allow a direct comparison with the EW(\hb) bias reported by \cite{Groves2012a}, we also compute the integrated residual fluxes normalized by the continuum level underneath the \hb\ line for individual spaxels. The EWs of the residual fluxes give a median value of 0.04 \AA, about one order of magnitude smaller than the value reported by \cite{Groves2012a}.

    In short, besides what \cite{Belfiore2019} have demonstrated on the robustness of the emission line measurements using mock spectra, we provide empirical evidence using the residual spectra from data to show that both the stellar continuum and the \ha\ and \hb\ lines are well modeled on average by the DAP. We thus believe that the flux measurements for \ha\ and \hb\ should be statistically unbiased at least for our main sample.
   
    \section{Results for other recombination lines}
    \label{appendix: res_otherlines}

    \begin{figure*}[htb!]
        \centering
        \includegraphics[width=0.9\textwidth]{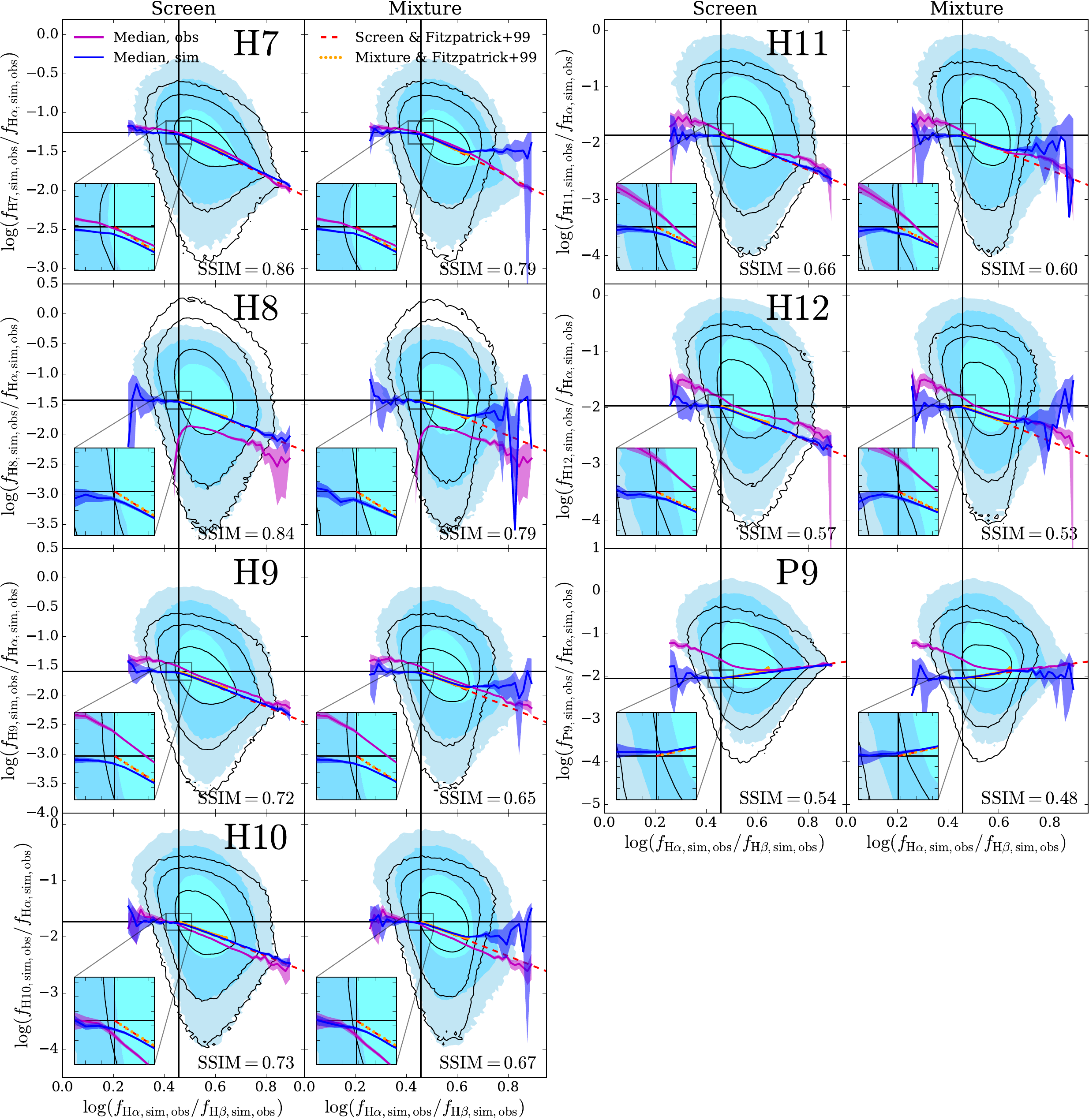}
        \caption{Similar to Fig. \ref{fig: lineratio_sim} but for other hydrogen recombination lines (i.e., H7, H8, H9, H10, H11, H12, and P9).
        \label{fig: lineratio_sim_others}}
    \end{figure*}

    \begin{figure*}[htb!]
        \centering
        \includegraphics[width=0.95\textwidth]{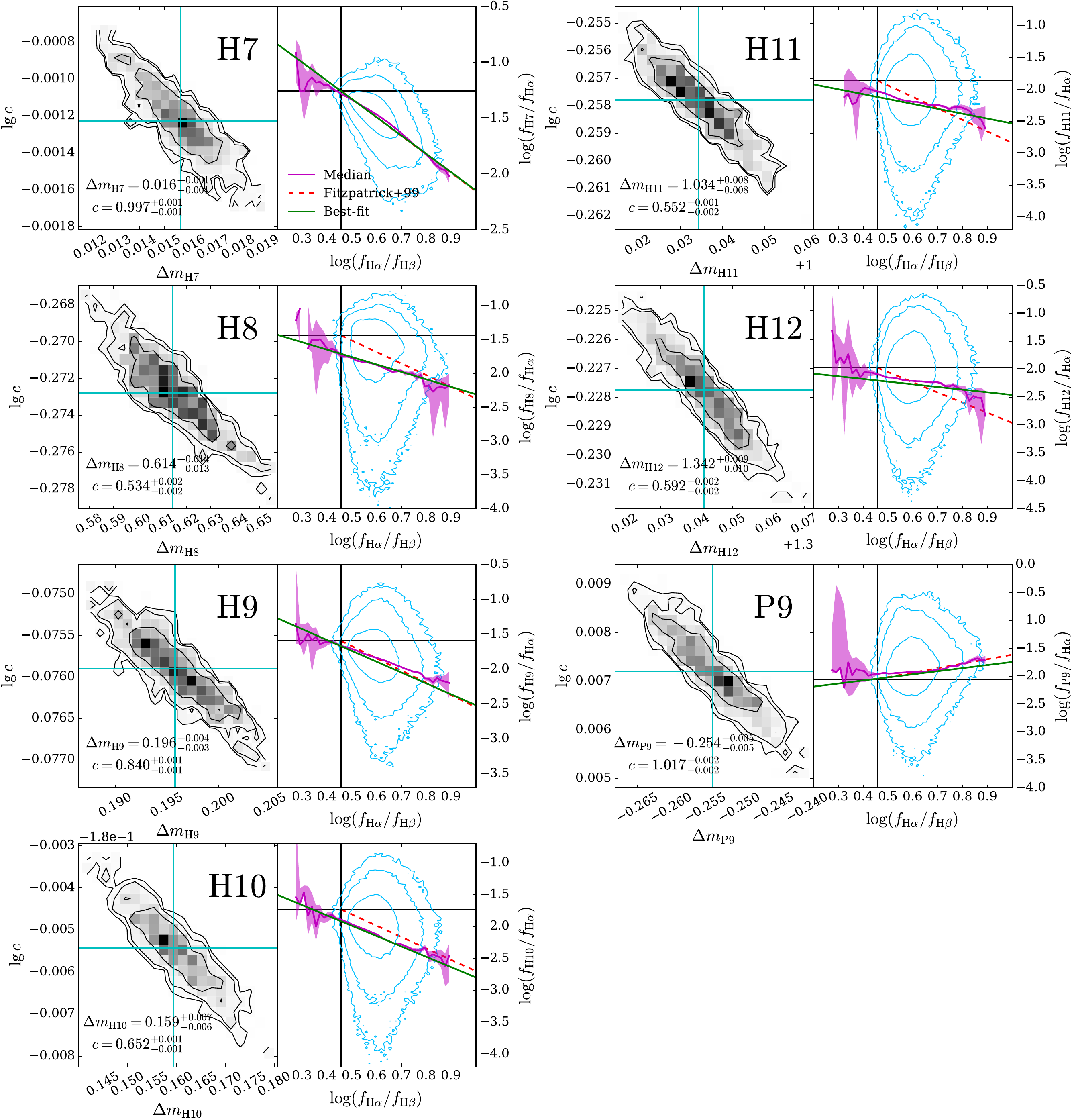}
        \caption{Similar to Fig. \ref{fig: res_minmc_bestfit} but for other hydrogen recombination lines (i.e., H7, H8, H9, H10, H11, H12, and P9).
        \label{fig: lineratio_bestfit_others}}
    \end{figure*}

    In the main part of this paper, we only present and discuss the results for the \hg, \hd, P8, and P10 lines, here we further provide results for the remaining hydrogen recombination lines that were measured by the DAP. In Figures \ref{fig: lineratio_sim_others} and \ref{fig: lineratio_bestfit_others} we provide the comparisons between the observed and simulated line ratio plots and the best-fit results for high-order Balmer lines (i.e., H7, H8, H9, H10, H11, and H12) and P9, respectively. Note that the median curves and the density contours of the observations are different in Figures \ref{fig: lineratio_sim_others} and \ref{fig: lineratio_bestfit_others} since different samples (the $\mathrm{SNR_{cut}}=3$ sample and the $\Sigma_{\rm SFR, cor}$-selected sample, respectively) are used to generate these plots (see Section \ref{subsubsec: simu_fitting_fitsubsample}). The best-fit parameters of these lines for the fitting sample are also given in Table \ref{tab: bestfit_paras} for reference.

    Similar to what we observe in Fig. \ref{fig: lineratio_sim}, the uncertainties of emission-line measurements exhibit prominent influences in the line ratio plots in terms of letting the median curves depart from the cross points predicted by the Case B recombination and scattering the line ratios to regimes that cannot be covered by the true values. When comparing the model predictions from the foreground screen and uniform mixture dust models with the observations, we can reach the same conclusion as Fig. \ref{fig: lineratio_sim}. Namely, most of the spaxels favor the foreground screen model rather than the uniform mixture one for the high-order Balmer lines up to H12 and P9. We inspect further for the foreground screen case and find more details that reveal the problem in the data quality.

    It seems that only the median curve of H7 can give a roughly consistent trend with the simulated one in terms of both slope and intercept. Although the difference in the intercept of the median curve between the observed and simulated (i.e., the predictions of the Case B recombination) samples of H7 are not as small as those of \hg\ and \hd, the best-fit $c$ is very close to 1 (see Fig. \ref{fig: lineratio_bestfit_others} and Table \ref{tab: bestfit_paras}), indicating a $<1\%$ correction for the model predicted fluxes. We thus believe that the slope of the attenuation curve at the wavelength of H7 can be well determined by our method.

    For H8, the observed median curve has a large systematic offset from the simulated one, which should be primarily attributed to the influence from the nearby \hei\ $\lambda$3888 line. These two lines can not be resolved from the MaNGA spectra, and thus their fluxes are very difficult to constrain via the DAP since no prior information about the fluxes is applied. From H9 to H12, all the observed median curves show deviations from the Case B predicted (simulated) ones to some extent. Similar results but for the $\Sigma_{\rm SFR, cor}$-selected sample can be found in Fig. \ref{fig: lineratio_bestfit_others}. The best-fit $c$ parameter of these high-order Balmer lines for the fitting sample ranges from 0.55 to 0.84, suggesting that substantial corrections should be applied to $f_{\rm X, pre}$ to match the observed fluxes.

    Sources of such systematic corrections might be the improper subtraction of the stellar continuum since the flux measurements of these Balmer lines should be corrected for evident stellar absorptions that strongly depend on the stellar continuum modeling. If true, our finding might imply that the stellar models involved in the DAP continuum modeling still need to be improved to recover the stellar absorptions of the Balmer lines more accurately. Correcting for stellar absorption becomes more critical for these high-order Balmer lines compared to those with smaller upper atomic level $n$ due to the dominant role of the stellar absorption over the nebular emission for these lines even for a very young stellar population \citep{GonzalezDelgado1999}. The weakness of these lines also implies that the zero-flux problem discussed in Section \ref{subsec: reasons_c_deviations} should have a large impact on the scaling factor.

    On the other hand, although we argue that the Case B recombination assumption works for the four lines we studied in the main text, departures from this assumption were also reported for some high-order recombination lines. \cite{MesaDelgado2009} observed enhanced intensities for both the Balmer and Paschen lines with an upper atomic level of $n\gtrsim 13$ compared to the Case B predictions in the Orion Nebula, while \cite{DominguezGuzman2022} further argued that for \hii\ regions in Magellanic Clouds, the deviations from the Case B values were found for recombination lines with $n>7$. However, the observational evidence relies on the adopted dust extinction curve (also see \citealt{DominguezGuzman2022}). Extinction-independent determination of the intrinsic intensities of these high-order lines for more \hii\ regions is required to confirm such deviations. If exist, the elevated intrinsic fluxes would result in $c>1$, which is in contrast with the best-fit results from H9 to H12 shown in Fig. \ref{fig: lineratio_bestfit_others}.

    For P9, the observed median curve given in Fig. \ref{fig: lineratio_sim_others} has a large upturn towards the low-\ha/\hb\ end, which can not be explained by the large uncertainties of emission-line measurements and implies a large overestimate of the P9 fluxes. Similar but less significant features have been observed for P8 and P10 displayed in Fig. \ref{fig: lineratio_sim}. We attribute this feature to the residuals of night-sky lines. It seems that the P9 line suffers more contaminations from these residuals compared to P8 and P10, which is consistent with our experience of visual inspection of individual spectra of spaxels with the Paschen line detections. The upturn feature is much weaker for the $\Sigma_{\rm SFR, cor}$-selected sample for which the best-fit $c$ is very close to 1, as shown in Fig. \ref{fig: lineratio_bestfit_others}. However, considering the influence of the sky residuals, we do not want to claim a good fitting for P9 in this work.

    In principle, adding the scaling factor as a free parameter in the fitting could ensure that $m_{\rm X}$ still can be well determined if the corrections to $f_{\rm X, pre}$ exist but are systematic and independent of the physical properties (more specific, the \ha-to-\hb\ ratio) of spaxels (e.g., a multiplicative systematic error of flux calibration). However, it is difficult to determine whether the main reason for the non-unity best-fit $c$ varies with the \ha-to-\hb\ ratio. Therefore, we claim that among these Balmer lines and P9, only H7 has a reliable fitting, and thus $m_{\rm H7}$ for the \hii\ region-dominated spaxels can be well determined.

    \section{Proof of an unchanged slope when using different fiducial curve}
    \label{appendix: proof_unchanged_mx}
    
    The definition of $m_{\rm X}$ (Equation \ref{eq: def_m}) contains the attenuation difference between \ha\ and \hb, which varies with the adopted extinction/attenuation curve. In Section \ref{subsec: fitting_method} we claim that the best-fit $m_{\rm X}$ should remain unchanged when using another fiducial curve in our fitting. To prove this statement, we here provide a brief derivation to demonstrate that for a given $m_{\rm X}$\footnote{Here we can treat this fixed slope as the best-fit one.} and a set of emission-line observations $f_{i,\rm obs}$ of one spaxel, our fitting approach, in principle, can return the same predicted flux $f_{i,\rm pre}$ for \hb\ and the targeted line and thus keep $\ln \mathcal{P}_{\rm spaxel}$ unchanged even when the attenuation difference $A_{\rm H\beta}-A_{\rm H\alpha}$ varies with the adopted fiducial curve. Notably, the above two statements are equivalent.
    
    We define a total-to-select attenuation between \ha\ and any other recombination line X as
    \begin{align}
        R_{\rm \alpha X} \equiv \frac{A_{\rm H\alpha}}{A_{\rm X}-A_{\rm H\alpha}},
    \end{align}
    then the relative slope $m_{\rm X}$ can be rewritten as
    \begin{align} \label{eq: mx_RR}
        m_{\rm X} = \frac{A_{\rm H\alpha}-A_{\rm X}}{A_{\rm H\beta}-A_{\rm H\alpha}}=-\frac{R_{\alpha\beta}}{R_{\rm \alpha X}}.
    \end{align}
    Assume that for a given $R_{\alpha\beta}$ and $R_{\rm\alpha X}$ (thus a fixed $m_{\rm X}$), the spaxel has a best-fit solution ($f_{\rm H\alpha,int}$, $A_{\rm H\alpha}$) that lets $\ln\mathcal{L}_{\rm spaxel}$ reach its maximum $\ln\mathcal{L}_{\rm spaxel, max}$. Below we show that if both $R_{\alpha\beta}$ and $R_{\rm \alpha X}$ are both scaled by a constant $\mathcal{C}$ (i.e., $m_{\rm X}$ remains unchanged), a new $A^{\prime}_{\rm H\alpha}=\mathcal{C} A_{\rm H\alpha}$ could let the $\ln\mathcal{L}_{\rm spaxel, max}$ unchanged (i.e., the model predicted fluxes $f_{i,\rm pre}$ in Equation \ref{eq: def_lnLsp} are unchanged).
    For H$\alpha$, requiring the predicted fluxes to be the same gives 
    \begin{equation}
        f_{\rm H\alpha, pre}=f_{\rm H\alpha, int} 10^{-0.4A_{\rm H\alpha}}=f^{\prime}_{\rm H\alpha, int} 10^{-0.4\mathcal{C}A_{\rm H\alpha}},
    \end{equation}
    thus the intrinsic \ha\ flux after the scaling should be $f^{\prime}_{\rm H\alpha, int}=f_{\rm H\alpha, int} 10^{0.4(\mathcal{C}-1)A_{\rm H\alpha}}$.
    Assume that the intrinsic line ratio between the recombination line X and \ha\ is $r_{\rm int, X\alpha}$, then the intrinsic flux of X after the scaling can be written as
    \begin{equation}
        f^{\prime}_{\rm X,int}=f^{\prime}_{\rm H\alpha,int} r_{\rm int, X\alpha}=f_{\rm X,int} 10^{0.4(\mathcal{C}-1)A_{\rm H\alpha}},
    \end{equation}
    in which $f_{\rm X,int}$ is the intrinsic flux of X before the scaling. Because $R_{\rm \alpha X}$ is scaled by $\mathcal{C}$, we have $R^{\prime}_{\rm \alpha X}=\mathcal{C} R_{\rm \alpha X}$ from which the new dust attenuation for line X can be obtained, i.e., $A^{\prime}_{\rm X}=A_{\rm X}+(\mathcal{C}-1)A_{\rm H\alpha}$. The predicted flux for line X after the scaling is
    \begin{align}
        f^{\prime}_{\rm X, pre}&=f^{\prime}_{\rm X, int} 10^{-0.4A^{\prime}_{\rm X}} \nonumber\\
        &=f_{\rm X, int} 10^{0.4(\mathcal{C}-1)A_{\rm H\alpha}}\times 10^{-0.4[A_{\rm X}+(\mathcal{C}-1)A_{\rm H\alpha}]}\\
        &=f_{\rm X, int} 10^{-0.4A_{\rm X}}\nonumber,
    \end{align}
    thus $f^{\prime}_{\rm X, pre}=f_{\rm X, pre}$. In other words, the predicted flux for line X remains unchanged after scaling $R_{\rm \alpha X}$ by a constant. Since line X can be any hydrogen recombination lines other than \ha, the relation between $m_{\rm X}$ and $R_{\rm \alpha X}$ (Equation \ref{eq: mx_RR}) further implies that for a given $m_{\rm X}$, a new best-fit solution ($f^{\prime}_{\rm H\alpha, int}$, $A^{\prime}_{\rm H\alpha}$) = ($f_{\rm H\alpha, int} 10^{0.4(\mathcal{C}-1)A_{\rm H\alpha}}$, $\mathcal{C}A_{\rm H\alpha}$) can return the same $\ln \mathcal{P}_{\rm spaxel, max}$ if both $R_{\alpha\beta}$ and $R_{\rm \alpha X}$ are scaled by a constant $\mathcal{C}$. Therefore, using another fiducial curve only changes $R_{\alpha\beta}$, but the best-fit $m_{\rm X}$ derived from our fitting method could remain in principle.
   
\end{appendix}

\end{document}